\documentclass[useAMS,usenatbib]{mn2e}
\usepackage{latexsym,amsmath,amssymb,enumerate,hyperref}
\usepackage[]{graphicx,lscape}
\usepackage{subfigure}
\usepackage{multirow}

\usepackage[perpage,symbol*]{footmisc}

\usepackage{verbatim}
\usepackage[usenames,dvipsnames]{color}

\usepackage{ulem}
\usepackage[usenames,dvipsnames]{color}

\def \jeans {\textsc{JEAnS} }
\def \jeansvii {t-\textsc{JEAnS} }
\def \mampost {{MAMPOSSt}}
\def \los {\mathrm{los}}
\def \vlos {v_\mathrm{los}}

\def \sigmar {\textcolor{black}{\sigma_{\mathrm{rr}}^2}}
\def \sigmat {\textcolor{black}{\sigma_{\mathrm{tt}}^2}}

\def \d {\mathrm{d}}

\def \rvir {r_{\mathrm{vir}}}

\def \sigmarcoeff {\textcolor{black}{c}_{\mathrm{r}}}
\def \sigmatcoeff {\textcolor{black}{c}_{\mathrm{t}}}
\def \KRlos {K^{\mathrm{Rlos}}}
\def \KTlos {K^{\mathrm{Tlos}}}
\def \Klos {K^{\mathrm{los}}}

\def \IRlos {\mathrm{I}^{\mathrm{r}}}
\def \ITlos {\mathrm{I}^{\mathrm{t}}}
\def \AICc {\mathrm{AICc}}
\def \BIC {\mathrm{BIC}} 

\def \DMS {\mathrm{DMS}}
\def \smoothLambda {\lambda}
\def \smoothBeta {\beta}

\def \GAIA {\textsc{Gaia Challenge }}

\def \knots {\boldsymbol{\xi}}

\def \sigmalos {\sigma_{\mathrm{los}}^2}

\def \bestsigmalos {\hat{\sigma}_{\mathrm{los}}^2}

\def \dtracer {\rho_{\star}}
\def \pdtracer {\Sigma_{\star}}

\def \dDM {\rho_{\bullet}}

\DeclareMathOperator{\diff}{d}
\DeclareMathOperator{\erf}{erf}

\newtheorem {theorem} {Theorem}

\renewcommand{\emph}{\textit}

\title[Reliable mass estimates]{Reliable mass calculation in spherical gravitating systems}
\author[Diakogiannis et al.]{Foivos I. Diakogiannis$^{1,2}$\thanks{E-mail:
foivos.diakogiannis@data61.csiro.au}, Geraint F. Lewis$^{3}$, 
Rodrigo A. Ibata$^{4}$, Magda Guglielmo$^{3}$,
\newauthor   
 Mark I. Wilkinson$^{5}$  and Chris Power$^{2}$\\
 $^{1}$Data61, CSIRO, Floreat WA, Australia\\
$^{2}$International Center for Radio Astronomy Research, University of Western Australia, 35 Stirling Highway, Crawley, WA 6009, Australia \\
$^{3}$Sydney
Institute for Astronomy, School of Physics, A28, University of Sydney, NSW 2006, Australia\\
$^{4}$Observatoire
Astronomique, Universit\'{e} de Strasbourg, CNRS, 11, rue de l Universit\'{e}, F-67000 Strasbourg, France\\
$^{5}$Department of Physics \& Astronomy, University of Leicester, Leicester LE1 7RH, UK}

\begin{document}



\maketitle

\label{firstpage}
\begin{abstract}
We present an innovative approach to the methodology of  dynamical modelling, allowing practical reconstruction of the underlying dark matter mass without assuming both the density and anisotropy functions. With this, the mass-anisotropy degeneracy is reduced to simple model inference, incorporating the uncertainties inherent with observational data, statistically circumventing the mass-anisotropy degeneracy in spherical collisionless systems. We also tackle the inadequacy that the Jeans method of moments has on small datasets,  with the aid of Generative Adversarial Networks: we leverage the power of artificial intelligence  to  reconstruct non-parametrically the projected line-of-sight velocity distribution.  We show with realistic numerical simulations of dwarf spheroidal galaxies that we can distinguish between competing dark matter distributions and recover the anisotropy and mass profile of the system. 
\end{abstract}


\section{Introduction}


Whilst dark matter represents the dominant mass component of the universe, its true nature remains elusive. Astrophysical probes of the properties of dark matter in large galaxies and galaxy clusters are typically hampered by the complexities of baryonic physics, and the complex coupling of the properties of kinematic tracers and the underlying form of the gravitational potential. 

In recent years, considerable focus has been given to dwarf spheroidal galaxies in the local universe. With a stellar mass of $\sim10^7 {\rm M_\odot}$, these are seen to be both devoid of gas, limiting the impact of baryonic astrophysics, and sufficiently simple to allow the determination of the gravitational potential of the dominant dark matter component from the stellar motions. However, traditional approaches of determining the distribution of dark matter in dSphs are limited by both the influence of the observational uncertainties and the mathematical complexity of deriving the properties of the dark matter. 

One such approach, the \citet{1979ApJ...232..236S} method, attempts to determine the underlying dark matter distribution through the reconstruction of the observed luminosity and kinematic properties of a galaxy using a library of precomputed orbits in trial potentials. Via the appropriate weighting of the components of the library for a particular mass model, the optimal fit to the data can be recovered and the mass determined. However, the computational aspects of the Schwarzschild method makes implementation highly impractical. Building a high-resolution orbit library to survey the likelihood of millions of mass models is currently computationally prohibitive. 

Other approaches are based upon the Jeans equation \citep{1980MNRAS.190..873B}, which relates the properties of kinematic tracers to the form of the gravitational potential. When applying the Jeans equation, there are two key ingredients, the distribution of dark matter, and a velocity anisotropy, $\beta$, which describes the relationship between radial and tangential orbits within the structure. In established approaches, it is typical to assume a functional form for the dark matter distribution, such as a Navarro-Frenk-White \citep{1996ApJ...462..563N} or a  \cite{1911MNRAS..71..460P} profile, and a functional form for $\beta$, optimizing the parameters of both based upon the observational data. Given the mathematical form of the Jeans equation, however, the resultant determination of the mass depends upon the assumed form for $\beta$, with various combinations of the adopted mass profile and $\beta$ providing equally acceptable fits to the data. Known as the ``mass-anisotropy degeneracy" (hereafter MAD), this is generally accepted as a fundamental limitation of Jeans-based approaches \citep[][see also \citealt{2017arXiv170104833R}]{1990AJ.....99.1548M}.

In this contribution, we present a new approach to address the MAD in the Jeans formalism, relying upon a parametrised functional form, known as a B-Spline, to account for the implicit relationship between the dark matter profile and the velocity anisotropy.
In this latest version of the \jeans \citep{2017MNRAS.470.2034D}    approach, the \emph{tight}-\jeans (hereafter t-\textsc{JEAnS}), we represent both the unknown radial    \emph{and} tangential  velocity dispersions as B-splines.  Then, we allow the data to give them the correct geometric shape. In this way, we avoid having to assume the functional form of all, but one, of the unknown functions used in the modelling process. 
Then, even with competing dark matter models that have equal numbers of unknown coefficients, we end up with statistical fits of different quality.  The key point is that by demanding that these curves be as  simple as possible, i.e. that they are represented by a minimal number of variables,  competing dark matter density models give different qualitative fits to the data. This eventually allows us to statistically discriminate between competing mass models and thus transform the MAD to a mere model inference problem. For the case of small datasets (of the order of ~1000 tracer stars), we use Generative Adversarial Networks (hereafter GANs, \citealt{NIPS2014_5423}) to reconstruct non-parametrically the underlying projected line-of-sight (LOS) velocity distribution. With this, we augment artificially the data to arbitrarily large numbers, and obtain reliable estimates for the moments of the LOS velocity distribution with an unprecedented density of points. The combination of \jeansvii modelling with the GANs for artificial data augmentation is a powerful approach for reliable mass estimates. 

In Section \ref{JEAnS_review} we present a short review of the Jeans mass modelling method. In Section \ref{JEAnS_data} we give the details of the datasets we used as well as the preprocessing method we followed. In Section \ref{JEAnS_section} we give a detailed description of the \jeansvii algorithm.  In Section \ref{section_results} we present our findings and in Section \ref{tJEAnS_discussion} we discuss the reasons behind the efficiency of the \jeansvii. Finally in Section \ref{section_conclusion} we present our concluding remarks.

\section{A review of the Jeans modelling methodology}
\label{JEAnS_review}

In this section we present an overview and analysis of  the established \citep{2008gady.book.....B} methodology of Jeans modelling.  We continue by providing a proof for the uniqueness of the anisotropy profile upon assuming a specific functional form for the  mass density profiles of stars, $\dtracer$, and dark matter, (hereafter DM), $\dDM$. 

The Jeans modelling approach subject to the assumption of spherical symmetry is fully contained in the following two equations: 
\begin{align}
\label{eq_Jeans_trad}
-\frac{d\Phi}{dr} &= \frac{1}{\rho_{\star}} \frac{d }{dr} \left( \rho_{\star} \sigmar \right) + \frac{2}{r} \beta(r) \sigmar \\
\label{eq_sigmaLOS_trad}
\sigmalos(R) &= \frac{2}{\pdtracer(R)} \int_R^{\rvir} \left(
1 - \beta(r) \frac{R^2}{r^2}
\right) \frac{r \rho_{\star} \sigmar }{\sqrt{r^2-R^2}} dr.
\end{align}
Here, $\Phi$ is the total potential of the system, $\dtracer$ the stellar tracer density, $\sigmar$ the radial velocity dispersion, $\pdtracer$ is the projected tracer surface density, $\sigmalos$ is the observed line-of-sight velocity dispersion,  $\beta$ is the anisotropy profile  defined by $
\beta(r) = 1 - \sigmat/ (2 \sigmar)$, and $R$ and $r$ are, respectively, the projected and 3D distance radii from the centre of the system. \textcolor{black}{Although the integral in Eq. \eqref{eq_sigmaLOS_trad} usually has infinity as its upper bound, here we  define $\rvir$ as the distance in which the DM mass density, $\dDM$, profile falls to approximately $\dDM(\rvir) \approx 200 \rho_{\mathrm{crit}}$. For all practical purposes, this is a useful numerical approximation that does not alter our findings.} 
With the exception of the observed LOS velocity dispersion, $\sigmalos$, and the projected tracer density profile, $\pdtracer$, all remaining functions ($\dtracer, \dDM,\sigmar,\beta$) are unknown and need to be determined from the data.  
Therefore, the system is underdetermined\footnote{One though needs to be precise in the definition of  the number of  ``unknowns''. Usually, we make assumptions for the functional form of these unknown functions that depend on some parameters. It is the number of these parameters  that define the necessary number of equations to close the system. Then, for either exact  (numerical solutions) or overdetermined systems (statistical fitting),  we evaluate each of the Equations \eqref{eq_Jeans_trad} and \eqref{eq_sigmaLOS_trad} in a set of distinct locations, $r_i$,  $R_j$ that are  equal or greater in numbers to the number of unknown parameters.}. 
In practice we can make a very good approximation to the functional form of  the tracer density profile, $\dtracer$ given deep photometry of the dSph, and we are thus left with three unknown functions, $\{\dDM(r),\sigmar(r),\beta(r)\}$ in a system of two equations.  

Although there are variations\footnote{These include, e.g. using higher moments \citep{2003MNRAS.343..401L} of $\sigmalos$, or different assumptions on the distribution function of the 
system, i.e. different penalty functions when comparing the LOS velocity dispersion with observables.} to the general methodology, the common  
established \citep{2008gady.book.....B}  starting point to solving this system of coupled integrodifferential equations with respect to the  unknowns $\dDM(r)$, $\beta(r)$ and $\sigmar(r)$,  is to assume  parametric functional forms for the DM mass density, $\dDM$, and the anisotropy  profile, $\beta(r)$. 
In an iterative approach (assuming for simplicity we have full knowledge of the tracer profile, $\dtracer$), one proposes a set of values for the parameters that define $\dDM$ and $\beta$, then solves the differential Equation \eqref{eq_Jeans_trad} with respect 
to\footnote{Subject to the boundary condition $\lim_{r\to \rvir} \sigmar \approx 0$.} $\sigmar$  and substitutes the result in Eq \eqref{eq_sigmaLOS_trad}. The validity of the numerical values of the parameters that define $\dDM$ and $\beta$ is tested by comparing  the model $\sigmalos$ with the observables. This iterative process is performed until some convergence criterion is met. 
  The rationale behind this approach is that when we consider parametric forms for $\dDM$ and $\beta$, the system becomes overdetermined (since Equations \eqref{eq_sigmaLOS_trad} and \eqref{eq_Jeans_trad} are evaluated in various distinct locations, $r_i$, $R_j$) and thus a solution exists. 

  It  needs to be emphasized though that once we make an assumption for the parametric form of one of the three unknown functions, $\{\dDM,\sigmar,\beta\}$, the system of two equations with (the remaining) two unknowns is closed. That is, the remaining two functions can be fully determined without the need for their parametric representation. \textcolor{black}{There exist published \citep{1982MNRAS.200..361B,1990A&A...234...93S,1992ApJ...391..531D,2010MNRAS.401.2433M}  exact solutions to the system of these equations (termed \emph{inversion} techniques) that make a parametric assumption for only one of the three unknown functions. These prove that it is an unnecessary assumption to assume two of the three unknown functions in parametric form. Usually, assuming more parametric forms than necessary, increases the uncertainty in the model parameters, thus making the distinction between competing mass models even more difficult\footnote{In addition, there are well known methods for solving numerically systems of coupled integrodifferential equations, such as finite differences and finite element methods \citep{solin2005partial,2011MNRAS.410.2003J},    wavelets \citep{bertoluzza2008numerical} and B-splines discretization \citep{hooollig2003finite}.} .}

It is insightful to separate the process of solving the  system of coupled integrodifferential Equations \eqref{eq_Jeans_trad} and \eqref{eq_sigmaLOS_trad}, in two distinct approaches: the exact numerical solution of the equations to perfect noiseless data and the statistical fitting to noisy data. Clearly, all conclusions we can draw from knowledge gained in exact solutions of the system of Jeans equations can be transferred to the case of statistical fitting, while the converse is not always true.  
In the following, we focus on the exact numerical solution.

\subsection{Uniqueness of the anisotropy profile for a given mass model}
In this section, we provide a theorem that upon making an assumption for the functional form of the tracer and DM mass densities, $\dtracer, \dDM$, and the LOS dispersion, $\sigmalos$  there exists a unique anisotropy profile, $\beta$.  For our purposes, we consider we have full knowledge of the above-mentioned  functions, $\dtracer, \dDM$ and $\sigmalos$. We  solve Eq \eqref{eq_Jeans_trad} with respect to $\beta(r)$ and substitute  it under the integral sign of Eq \eqref{eq_sigmaLOS_trad}. Then we end up with a single integrodifferential  equation
 (subject to the virial boundary condition $\lim_{r\to \rvir} \sigmar (r) \approx 0$), namely
\begin{multline}
\label{eq_sigmaLOS_beta_free}
\sigmalos(R) = \frac{2}{\pdtracer(R)} \int_R^{\rvir} 
\biggl[ K_A \biggl( \frac{\diff (\dtracer  \sigmar )}{\diff r} + \dtracer  \frac{\diff \Phi}{\diff r} \biggr)  \\
+K_B 
\rho_{\star} \sigmar  \biggr] \diff r 
\end{multline}
where $K_A$ and $K_B$ are kernel functions defined by: 
\begin{align*}
K_A(r,R) &= \frac{R^2}{\sqrt{r^2-R^2}}, & 
K_B(r,R) &= \frac{2 r}{\sqrt{r^2-R^2}} 
\end{align*}
This equation has one unknown,  the radial velocity dispersion,  $\sigmar$. That is, assuming perfect knowledge of the   LOS velocity dispersion profile, $\sigmalos(R)$,   if we could solve this equation for the unknown $\sigmar$ we would obtain for each assumption of a mass model, $\{ \dtracer, \Phi(r) \}$, a radial velocity dispersion, $\sigmar$.  The question arises: is the solution with respect to $\sigmar$ unique? 
\begin{theorem}
\label{JEAnS_theorem}
The solution of Eq. \ref{eq_sigmaLOS_beta_free} with respect to $\sigmar$ for given $\sigmalos(R)$, $\dtracer (r)$ and $\Phi(r)$ profiles, is unique. 
\end{theorem}
We provide the proof of Theorem \ref{JEAnS_theorem} in Appendix \ref{Theorem_1_proof}.   
\textcolor{black}{This result, which complements the published inversion techniques, has the following implication: we can assume only one of the three unknown functions and thus reduce \emph{the uncertainty} of the modelling parameters (in comparison with the uncertainty we get by assuming parametric forms for two unknown functions, as is customary). For the case of statistical fitting, we can use hierarchical models (e.g. smoothing splines) of varying complexity, that make model selection possible and this is the key for breaking \emph{statistically} the Jeans degeneracy: the missing ingredient (an additional equation) is replaced by the model selection criterion. 
Thus, if we allow the total mass, $M(r)$ to vary, i.e. if we assume a different mass model for the same $\sigmalos$ profile, then the anisotropy profile will generally be different. However there is one important constraint we need to consider, namely, the projected virial theorem (discussed in detail in Section~\ref{sec:Projected_virial_theorem}). The projected virial theorem does not depend on the anisotropy, which implies that not all mass profiles are consistent with the projected kinetic energy evaluated from $\sigmalos$. However, the projected virial theorem, on its own, is not sufficient to break the degeneracy (it is a single scalar equation, therefore the total number of equations is still less than the unknowns). It can only further reduce the feasible solution space of where the $M(r)$ function resides.}
We will discuss this further in Section  \ref{tJEAnS_discussion}. In Section \ref{section_results_exact} we provide numerical examples of the uniqueness of the kinematic profile for an assumed mass density.

It should be stated that we can choose equally well to assume a functional form for the anisotropy profile, and leave the mass density to be deduced by the data \citep[][see also \citealt{2017arXiv170104833R}]{2010MNRAS.401.2433M}: in this case, the total mass of the system \emph{follows} from the assumptions of the anisotropy model, $\beta$, for a given data set of observables. 
This can be very easily seen from the following: again,  assuming perfect knowledge of $\sigma_{\mathrm{los}}^2$ profile, once we use a specific functional form for the anisotropy 
$\beta$, the system of equations that describes a stellar dynamical system is: 
\begin{align}
\label{DLI_nat_beta_ass_1}
\frac{1}{\pdtracer}
\int_R^{\rvir} \biggl[ \rho_{\star} K_1(r,R) \sigmar +  \rho_{\star} K_2(r,R) \sigmat \biggr] \diff r
&= \sigmalos(R) 
\\
\label{DLI_nat_beta_ass_2}
\sigmat -2(1-\beta(r)) \sigmar  &=0 \\ 
\label{DLI_nat_beta_ass_3}
 \frac{1}{\rho_{\star}} \frac{d }{dr} \left( 
 \rho_{\star}
 \sigmar \right) 
 + \frac{2 \sigmar-\sigmat}{r} 
&=  -\frac{G M_{\mathrm{tot}}(r)}{r^2} 
\end{align}
These are three equations with respect to the three  unknowns $\sigmar, \sigmat$ and the total mass $M_\mathrm{tot}$. The system of equations, Eq.   \ref{DLI_nat_beta_ass_1}, \ref{DLI_nat_beta_ass_2} is complete, i.e. we have a unique solution for $\sigmar$ and $\sigmat$.  Then, from the last Eq. \ref{DLI_nat_beta_ass_3} we calculate the total mass, $M_{\mathrm{tot}}$, whose value depends solely on the tracer stellar density, $\dtracer$, and the kinematic profile, $\sigmar, \sigmat$. 
 Therefore, when we model  by \textit{assuming} a specific anisotropy profile, $\beta$,  we effectively pre-specify the mass content of the system. In this reasoning, we did not need to adopt any assumptions for the parametric form of the DM mass density.

\section{Data}
\label{JEAnS_data} 
In this section we provide an overview of the datasets we used to validate our methodology. \textcolor{black}{We describe how we pre-process the data and create validation and test data sets for the \jeansvii solver, as well as how we use GANs to generate large artificial samples of data for the case of small datasets}.

We test our algorithm with  the \GAIA\footnote{http://astrowiki.ph.surrey.ac.uk/dokuwiki/doku.php} suite of mock simulations, in particular the spherically symmetric sets. The mock suites provide a snapshot of the full 6D information of the tracer profile, $x,y,z,v_x,v_y,v_z$. 
For modelling each of the systems, we used only the projected positions, $x,y$ and the LOS velocity, $v_z$. Following the \GAIA  guidelines, we used the data that include velocity errors. For each datum we considered that this error is equal to the $2\%$ of the true $v_z$ velocity.
Our training data set, $D_{\mathrm{train}}$, consists of the values, $\{x,y,v_z\}_j$, $j=1,\ldots,N_{\star}$, as well as the second order  moments, $\sigma_{\mathrm{los}}^i$, of the projected LOS velocity. 

As a  sample of the various datasets, we chose  the PlumCuspOM, PlumCuspIso, PlumCuspTan and NonPlumCoreOM suites. \textcolor{black}{For the first three we use the suites with 10k targets, and for the last one we use both 10k and 1k datasets.}  The Plummer-like family of tracer profiles was chosen based on the knowledge that most Stellar profiles observed in nature are cored; the latter three models were considered to be representative by the curator of the \GAIA \citep{2017arXiv170104833R}. \textcolor{black}{In particular, the NonPlumCoreOM is a notoriously difficult set to model, and this is the reason why for this particular one we also include a set with only 1k targets.} 
Each of these datasets was modelled with assumptions for the stellar and DM profiles only. We model each system with two competing models: one with the true parametric form (with parameters recovered from the fitting process), and one with an incorrect parametric assumption (again the parameters are fitted to the data). \textcolor{black}{We report the combinations of Stellar and DM models we used in Tables \ref{JEAnS_test_models_2} and \ref{JEAnS_test_models_2_1k}.} In all model fits, the anisotropy profile is evaluated from the data.   

The reference anisotropy profiles that these data sets were created from are two, namely constant and 
Ossipkov-Merritt \citep{1979SvAL....5...42O,1985AJ.....90.1027M}:  
\begin{equation}
\beta(r) = 
\begin{cases}
 \beta_0, &
\mathrm{constant}\\
 \dfrac{r^2}{r^2+r_{\mathrm{a}}^{\alpha}}, & \text{OM}
\end{cases}
\end{equation}
The  mass density profile that these data sets  follow, both for the stellar and the DM components, is given by a power law form \cite{1996MNRAS.278..488Z}, for a variety of reference parameters:
\begin{equation}\label{GAIA_mass_profile}
\rho(r) = \rho_0 \left(\dfrac{r}{r_{\mathrm{s}}}\right)^{-\gamma} 
\left[
1+ \left(
\dfrac{r}{r_{\mathrm{s}}}
\right)^{\alpha}
\right]^{(\gamma-\beta)/\alpha}
\end{equation}
In addition, the \GAIA datasets make the approximation that the stellar tracer mass is negligible in comparison to the DM mass component.
In order to account for this we normalized the total tracer mass to unity, i.e. $M_{\star}^{\mathrm{tot}} = 1M_{\odot}$.

We report the reference model  parameters for each of the datasets used in Table \ref{JEAnS_test_table_1}. \textcolor{black}{ In Tables \ref{JEAnS_test_models_2} and \ref{JEAnS_test_models_2_1k} we record the combinations of Stellar and DM mass models we used for the modelling process as well as the  test error for the various competing models.}

\begin{table*}
\caption{\textcolor{black}{Synthetic data sets: parameters $\alpha,\beta,\gamma$ are dimensionless numbers. Distance parameters $r_{\star,\bullet}$ are in kpc, while $\rho_{0\bullet}$ in $\mathrm{M}_{\odot} \mathrm{pc}^{-3}$.}
}
\label{JEAnS_test_table_1}
\begin{center}
\begin{tabular}{ | l |   c | c | c | }\hline
DataSet   & $\theta_{\star} = [ r_{\star}, \alpha_{\star}, \beta_{\star}, \gamma_{\star} ]$ & $\theta_{\bullet} = [ \rho_{0\bullet},  r_{\bullet}, \alpha_{\bullet}, \beta_{\bullet}, \gamma_{\bullet} ]$ & $\beta(r)$ anisotropy   \\ \hline \hline
PlumCuspOM (10k)  &  $[0.1,2,5,0.1]$  &  $[0.064,1,1,3,1]$   & $\beta_{\mathrm{OM}}$, $r_a=0.1$  \\ 
\textcolor{black}{PlumCuspIso (10k)} & \textcolor{black}{$[0.25,2,5,0.1]$} &
\textcolor{black}{$[0.064,1,1,3,1]$} & \textcolor{black}{$\beta_0 = 0.0$}  \\ 
\textcolor{black}{PlumCuspTan (10k)} & \textcolor{black}{$[0.5,2,5,0.1]$} & \textcolor{black}{$[0.0239, 2 , 1,4,1]$}   & \textcolor{black}{$\beta_0= -0.5$}  \\ 
\textcolor{black}{NonPlumCoreOM \textcolor{black}{(1k, 10k)}} & \textcolor{black}{$[0.25,2,5,1]$} &  \textcolor{black}{$[0.400,1,1,3,0]$} & \textcolor{black}{$\beta_{\mathrm{OM}}$}, \textcolor{black}{$r_a=0.25$} \\ 
\hline\hline
\end{tabular}
\end{center}
\end{table*}

\begin{table}
\caption{\textcolor{black}{Competing mass models for the various 10k data sets. We report the  average error on unseen test data, $D_{\mathrm{test}}$. The  true models from which the data were produced are with  {\bf bold} fonts. In all cases the test error, $\chi^2_{\mathrm{test}}$, selects the correct model.}}
\label{JEAnS_test_models_2}
\begin{center}
\begin{tabular}{ | l |  c | c | c | c | }\hline
DataSet   & Stellar model & DM model      & $\chi^2_{\mathrm{test}}$ \\ \hline \hline
PlumCuspOM, 10k & Plummer & Burkert &    223.972 \\
PlumCuspOM, 10k & {\bf gH}  &  {\bf NFW }   & \textbf{217.337} \\
\hline \hline
PlumCuspIso, 10k  & Plummer & Burkert   & 207.001   \\
PlumCuspIso, 10k  & {\bf gH} & {\bf NFW}    & {\bf 204.812} \\
  \hline \hline
PlumCuspTan, 10k  & Plummer & Burkert  &  192.115 \\
PlumCuspTan, 10k  & {\bf gH} & {\bf gH}  &  \bf{192.017} \\\hline \hline
NonPlumCoreOM, 10k  & Plummer & NFW  & 348.126  \\
 NonPlumCoreOM, 10k  & {\bf  gH} & {\bf  gH }   & {\bf 340.255} \\
\hline
\end{tabular}
\end{center}
\end{table}

\begin{figure}
\centering
\includegraphics[width=\columnwidth]{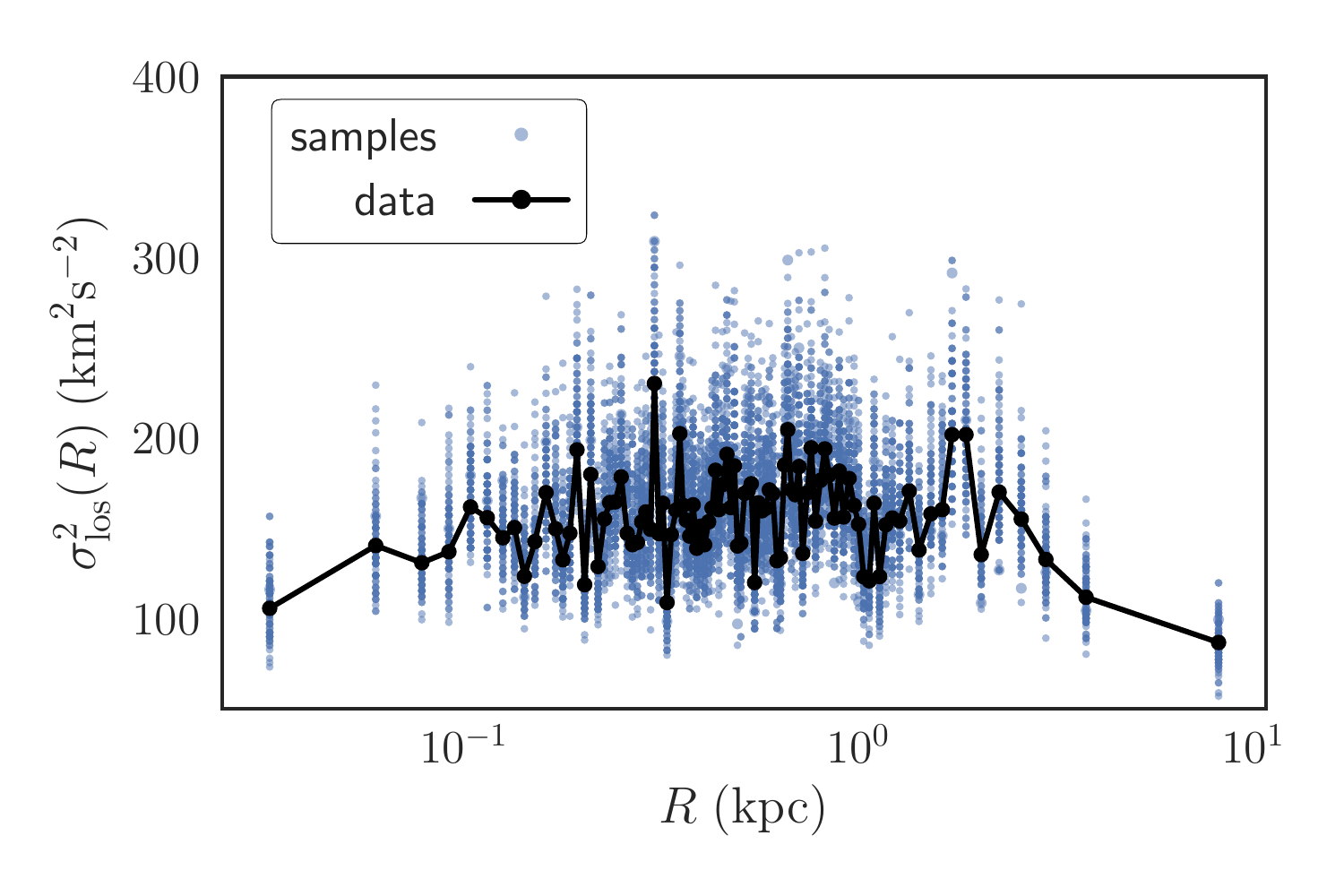}
\caption{ Binned $\sigmalos$ (black solid line) data as well as validation $\sigmalos$ dataset $D_{\mathrm{val}}$ for the PlumCuspTan model.}
\label{DLI_PlumCuspTan_DValidation}
\end{figure}

\begin{figure*}
\centering
\includegraphics[width=\textwidth]{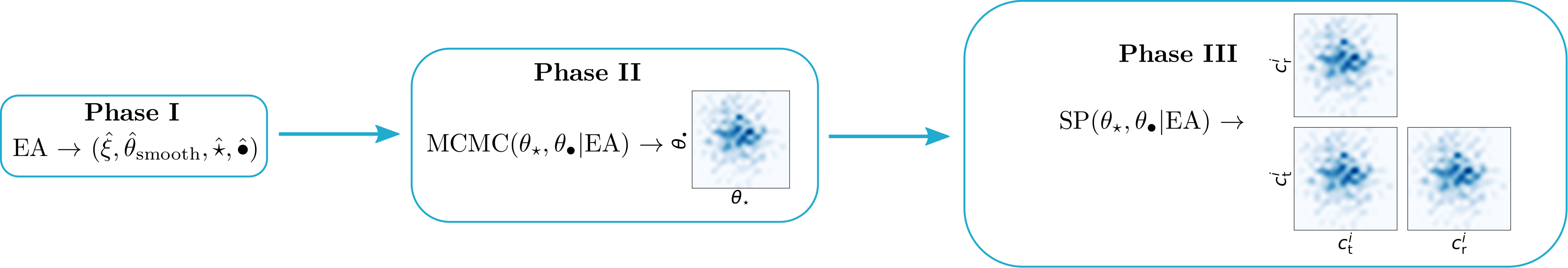}
\caption{ The \jeansvii algorithm.}
\label{DLI_nat_JEAnS_diag}
\end{figure*}

\subsection{Data preprocessing}
\label{thJm_data_prep}

In this section we describe the process that we followed in order to create the LOS velocity dispersion, $\sigma^i_{\mathrm{los}}$, values, that we use as training data.

Each bin contains $N_{\mathrm{targets}} = \sqrt{N_{\star}}$. For example, for $10^4$ stars, we have $N_{\mathrm{bin}}=10^2$. This approach gives equal Poisson error
 \citep{2017arXiv170104833R} for each datum. We modelled the distribution of stars, within each $i$ bin, as a Gaussian centred at zero. The likelihood of this model,  for each bin $i$, is:
\begin{equation}
\label{tgJm_bin_loglkhood}
P_i(v_j|s,D) = \prod_{j=1}^{N_{\mathrm{targets}}}
\frac{\exp \{ -v_j^2  (s^2 + (\delta v_j)^2)^{-1}/2 \} }{\sqrt{2 \pi (s^2+(\delta v_j)^2)}}.
\end{equation}
where $v_j$ is the value of the LOS velocity of star $j$ in bin $i$, $\delta v_j$ is the associated error, and $s$ the standard deviation of the Gaussian distribution. 
For each bin $i$, we perform an MCMC process using the likelihood Eq \eqref{tgJm_bin_loglkhood}, to estimate the marginalized distribution of the parameter $s$. 
It should be clear that this MCMC process is used only in the data pre-processing stage. It should not be confused with the MCMC we perform later for the estimation of marginalized distributions for the stellar, $\theta_{\star}$, and DM, $\theta_{\bullet}$, parameters. 
The LOS velocity dispersion data values, $\sigma_{\mathrm{los}}^i$, we use  at each location $R_i$ (centre of the $i$th radial bin), is the mode value of the histogram, 
$\hat{s}^2 = \sigma_{\mathrm{los}}^i$. The associated error, $\delta \sigma_{\mathrm{los}}^i$,  is the 1$\sigma$ uncertainty of $s^2$. 
Thus, $D_{\mathrm{train}}=\{x,y,v_z \}_j \cup 
\{\sigma_{\mathrm{los}}^i, \delta \sigma_{\mathrm{los}}^i \}$, $j=1,\ldots,N_{\star}$ and $i=1,\ldots,N_{\mathrm{bin}}$.

 In addition to the above LOS moments, we draw 100 random samples,  $\sigma_{\mathrm{los}}^{ij}$ ($j=1,\ldots,100$),  from the marginalized distribution of $\sigma_i^2$, that we keep for a validation data set, $D_{\mathrm{val}}$, and 200 random samples that we use for test sets, $D_{\mathrm{test}}$. During the Evolutionary Algorithm  (hereafter EA)  training, the validation set is used for the selection of the smoothing parameters, $\theta_{\mathrm{smooth}}$ (Section \ref{JEAnS_section}). During the MCMC training phase, 
instead of using the mode value $\sigma_{\mathrm{los}}^i$  of the LOS dispersion as the moments data, in each iteration of the solver, we select  random realizations, $\sigma_{\mathrm{los}}^i = \sigma_{\mathrm{los}}^{ij}$ (random $j$),  from the validation data set, $D_{\mathrm{val}}$. In this way we incorporate the uncertainty of the moments data as prior information to the modelling process.
The test set, $D_{\mathrm{test}}$, is used for the model selection between competing models after the EA phase. In Fig. \ref{DLI_PlumCuspTan_DValidation} we plot for the case of the PlumCuspTan model the binned $\sigma_{\mathrm{los}}^i$ values (black solid line), as well as the validation values, $\sigma_{\mathrm{los}}^{ij} \in D_{\mathrm{val}}$, for each bin $i$.  

\subsection{Data augmentation for small datasets using GANs}

In this section we briefly describe the application of GANs  for the numerical reconstruction of the 3D projected LOS velocity distribution, $f(x,y,\vlos)$ from the NonPlumCoreOM 1k dataset. Our goal is  to give an intuitive understanding behind the reason that this method is so effective and not to detail the GAN methodology (see \citealt{DBLP:journals/corr/Goodfellow17}  for a pedagogical introduction).

A fundamental limitation to the method of moments, in the Jeans framework, is that it requires a wealth of data to be successful. This is because the moments of the data, as a product of the summary information of the underlying distribution, are much fewer in number than the original unbinned dataset. This is more evident especially when the original dataset is small (from few hundred to 1k stars)  as is often the case in astronomical datasets (e.g. of dSph galaxies).  We overcome this difficulty by applying a preprocessing step, where we create synthetic data from a generative model, that resembles the true underlying distribution. That is, we create synthetic data to complement the original dataset and thus acquire a large number of LOS velocity moments. We do so only for the 1k NonPlumCoreOM dataset (although the method can be applied to the 10k as well for higher quality results). 
For this task, artificial intelligence actors (GANs) are excellent generative models, since they learn by ``looking'' at the real data, i.e. \emph{by example}, and are not bound by assumptions of the mathematical form of the underlying distribution.  

The general framework of the GANs consists of a set of two competing artificial neural networks (hereafter ANNs). The first, the Generator (hereafter $G$), takes as input a vector of random numbers and tries to create fake (synthetic) data whose distribution resembles  the distribution of the true training dataset. The second, the Discriminator (hereafter $D$), takes as input, true data, drawn randomly from the training distribution, or fake data, created randomly from $G$, and tries to predict whether the data that it was given are genuine (real) or fake. During training, the goal of $G$ is to make $D$ perform a mistake, i.e. the goal of $G$ is to generate as authentic looking synthetic data as possible. The goal of $D$ is to discriminate the true data from the fake ones and debunk the efforts of $G$. This framework is a minimax two-player game. During training both players become proficient in their task. When this process reaches equilibrium, $G$ is a faithful  approximator of the true underlying distribution of the training dataset. This method is unsupervised training which in practice means there is no upper bound on the quality of the data approximation. 

This method has been applied successfully, with impressive results, in artificial intelligence generative tasks, such as  the creation of high quality  images \citep{DBLP:journals/corr/abs-1710-10196}, for the creation of synthetic MRI scans for enhanced deep neural network training \citep{2018arXiv180710225S}, for motion transfer in videos \citep{2018arXiv180807371C} and many more cases where the data distribution is anything but ``easy'' to express mathematically (if not impossible). 

For our particular needs we construct a \textsc{pytorch} \citep{paszke2017automatic} implementation of  Wasserstein GANs with gradient penalty (hereafter WGAN-GP, \citealt{DBLP:journals/corr/GulrajaniAADC17}). We chose WGAN-GP because it is one of the most reliable GAN frameworks for stability in training. 
The architectures we used for the $G$ and $D$ ANNs are summarized in Table \ref{JEAnS_GANs}. The input to the generator is a random 10 dimensional multinomial distribution, $z \sim \mathcal{N}(0,1)^{10}\in \Re^{10}$. In Table \ref{JEAnS_GANs_params} we detail the hyper parameter values we used during GANs training. 
 In addition, in order to avoid overfitting the NonPlumCoreOM 1k data set, we augmented the data  with random rotations on the $x,y$ plane and reflections with respect to $x$ and $y$ axis. In particular we followed the transformations $(x,y,\vlos) \to (-x,y,-\vlos)$ and $(x,y,\vlos) \to (x,-y,-\vlos)$. For zero mean $\vlos$ stellar systems, these reflections are like observing the target from the opposite direction of the initial observer: clearly the physics of the system should not change. This type of information should be viewed as ``prior knowledge encoding'' of the modelling process with neural networks. 

\begin{figure}
\centering
\includegraphics[width=\columnwidth]{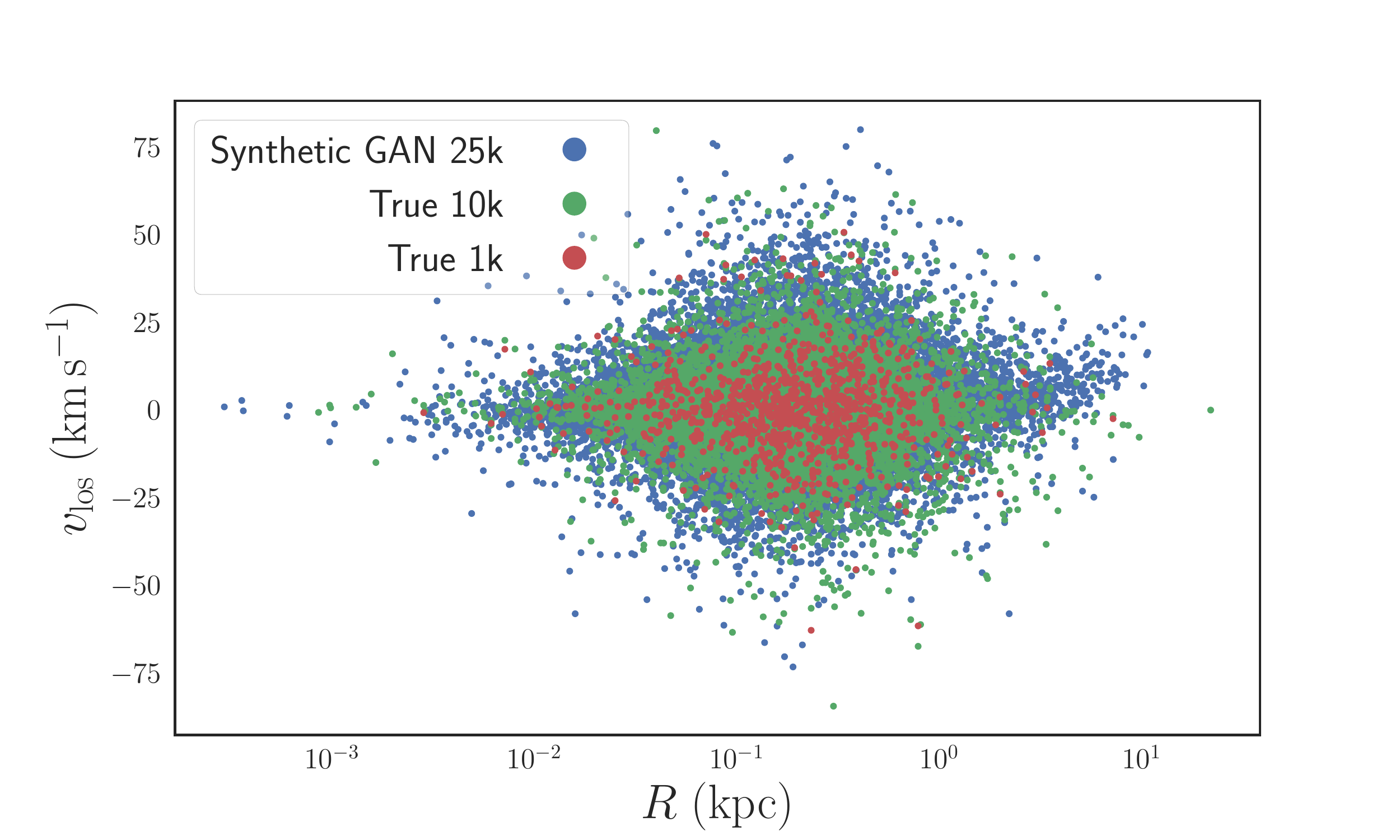}
\caption{ Comparison of  NonPlumCoreOM 1k, 10k true datasets against the GAN generated synthetic data on the $(R,\vlos)$ plane. }
\label{DLI_NonPlumCoreOM_True_vs_GAN_scatter2}
\end{figure}

\begin{figure}
\centering
\includegraphics[width=\columnwidth]{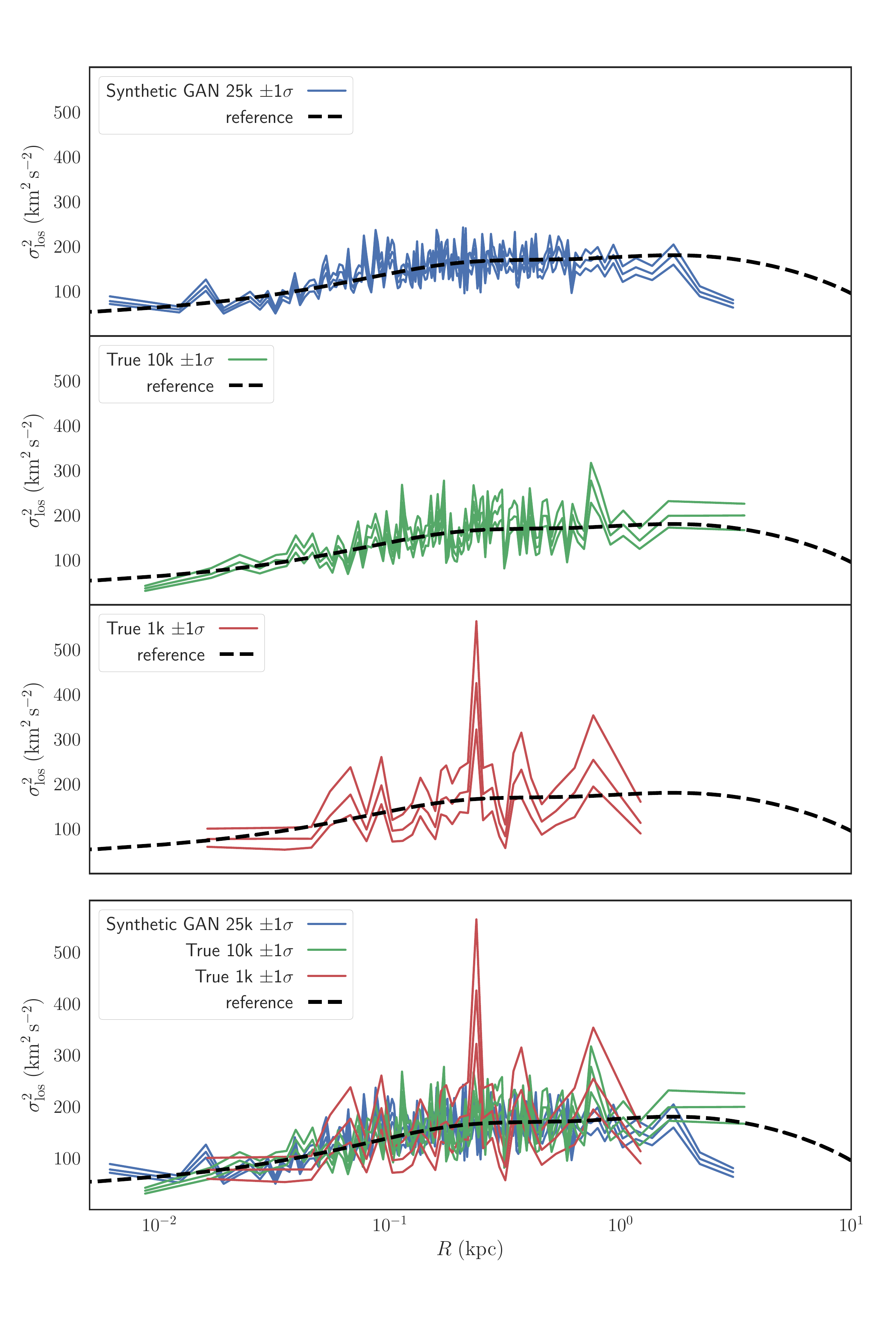}
\caption{ Comparison of  NonPlumCoreOM 1k, 10k true dispersion profiles against the GAN generated dispersion profile. Overplotted is the true reference profile.  The GAN generated profile was created from information from the NonPlumCoreOM 1k dataset only.}
\label{DLI_NonPlumCoreOM_True_vs_GAN_sigmaLOS}
\end{figure}

\begin{figure}
\centering
\includegraphics[width=\columnwidth]{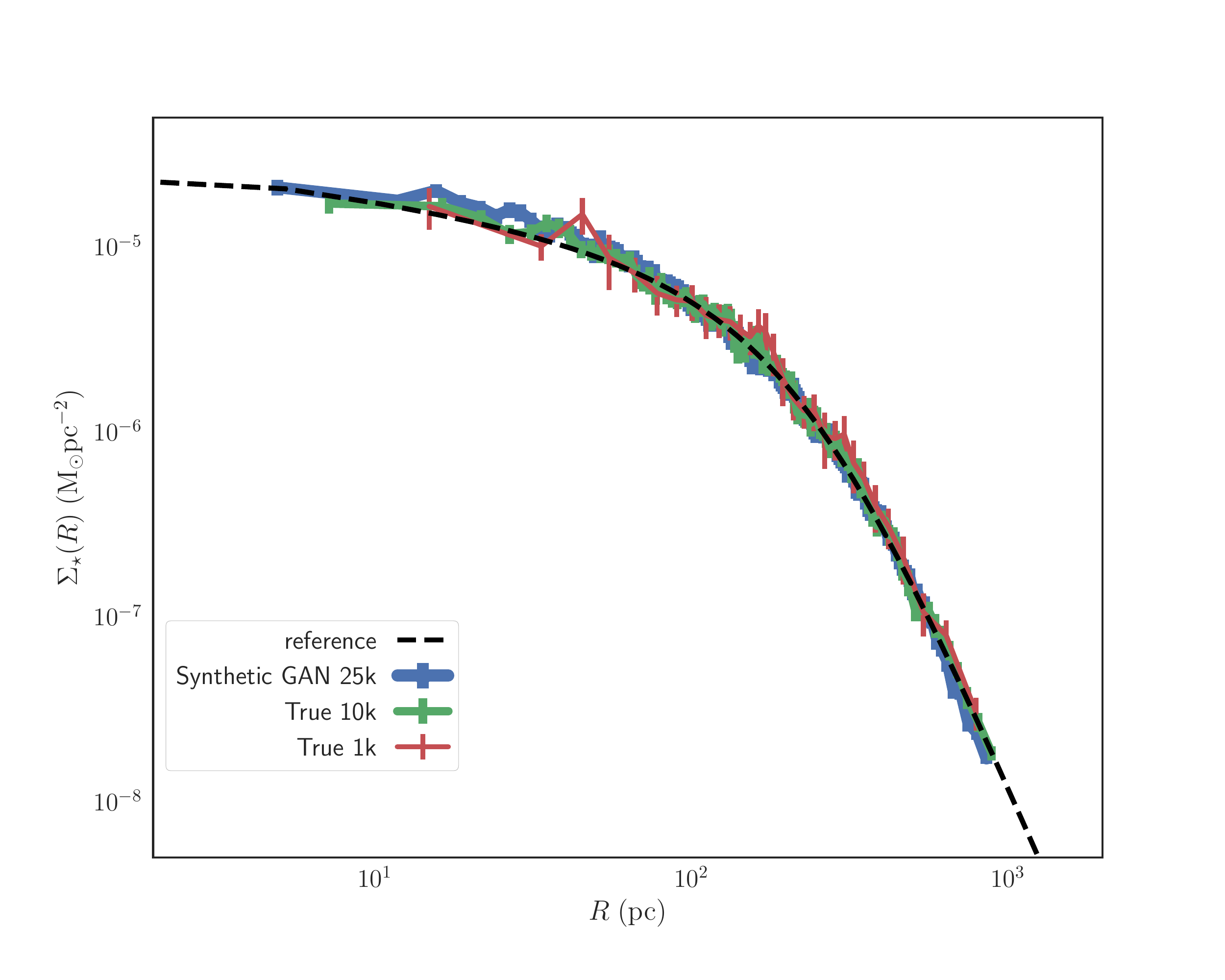}
\caption{ Comparison of  NonPlumCoreOM 1k, 10k true projected density profiles against the GAN generated projected density profile. Overplotted is the true reference profile.  }
\label{DLI_NonPlumCoreOM_True_vs_GAN_bright}
\end{figure}

In Fig \ref{DLI_NonPlumCoreOM_True_vs_GAN_scatter2} we plot on the $(R,\vlos)$ plane the synthetic data generated from the GANs  against the 1k and 10k NonPlumCoreOM datasets. We generated $\sim$25k synthetic data points by training the Discriminator, $D$, on the NonPlumCoreOM 1k dataset. This resulted in approximately 160 $\sigmalos$ binned values for the LOS velocity dispersion profile. 
In Fig. \ref{DLI_NonPlumCoreOM_True_vs_GAN_sigmaLOS} we plot the LOS velocity dispersion profile from the GAN data as well as the true 1k and 10k dispersion profiles. In all panels the reference profile (dashed curve) is overplotted. Clearly, the GAN generated profile is of high quality. In fact, the uncertainty of the data points around the reference profile is smaller than even the case of the original 10k dataset. This happens because the GAN system learns more information of the underlying distribution from the  NonPlumCoreOM 1k dataset than what the moments of the 10k sample can describe. As a result, with higher number of targets (25k) we end up with a LOS velocity dispersion profile of smaller uncertainty than the 10k original dataset. 
A small bias is apparent in the last two $\sigmalos$ data points, probably because the GANs overfit the outliers at the edges of the radial distance of the 1k dataset. This bias may also be due to the system of GANs not having reached the optimum equilibrium when we terminated training. Finally in Fig \ref{DLI_NonPlumCoreOM_True_vs_GAN_bright} we compare the projected density (brightness for $\Upsilon_{\mathrm{V}} = 1$) of the tracer population.
It should be noted that we did not experiment with new architectures, training schemes or hyperparameter optimization. We just used the proposed implementation scheme from \cite{DBLP:journals/corr/GulrajaniAADC17} for their toy model of 25 2D Gaussian distributions. There is huge scope for improvement and adaptation for individual datasets of this technique for data augmentation in astronomy in various sub-disciplines. Here, we are merely scratching the surface of the potential of this technology.

\begin{table}
\caption{Generator and Discriminator network architectures. We follow \textsc{pytorch} semantics to denote the dimensionality and type of the layers and non-linear activations we used. Here \texttt{LDIM}=10 is the dimensionality of the latent space that we sample and feed into the Generator, \texttt{DIM}=512 is the number of features in the linear Layers and \texttt{XDIM}=$\dim (x,y,\vlos)=3$ is the dimensionality of the projected observation space. }
\label{JEAnS_GANs}
\begin{center}
\begin{tabular}{ | l |  l | l | }\hline
Layer  & Generator& Discriminator \\ \hline \hline
 1& \texttt{Linear(LDIM,DIM)} & \texttt{Linear(XDIM,DIM)} \\\hline
 Activation& \texttt{LeakyReLU($\alpha=0.01$)} & \texttt{LeakyReLU($\alpha=0.01$)}\\\hline
2& \texttt{Linear(DIM, DIM)}& \texttt{Linear(DIM, DIM)} \\\hline
Activation& \texttt{LeakyReLU($\alpha=0.01$)} & \texttt{LeakyReLU($\alpha=0.01$)} \\\hline 
3& \texttt{Linear(DIM, DIM)} & \texttt{Linear(DIM, DIM)} \\\hline
  Activation& \texttt{LeakyReLU($\alpha=0.01$)} & \texttt{LeakyReLU($\alpha=0.01$)} \\\hline
4& \texttt{Linear(DIM, XDIM)} &  \texttt{Linear(DIM, 1)} \\\hline
\hline
\end{tabular}
\end{center}
\end{table}

\begin{table}
\caption{Training hyper parameters of the GANs system. \texttt{NBATCH} is the batch size, \texttt{NCRITIC} is the number of training iterations the $D$ performs for a single $G$ training iteration, \texttt{LDIM} is the dimensionality of the input random number to the $G$. For the gradient descent we used the Adam  optimizer \citep{DBLP:journals/corr/KingmaB14}. The input dataset that the $D$ was trained on was the NonPlumCoreOM 1k. }
\label{JEAnS_GANs_params}
\begin{center}
\begin{tabular}{ | l |  c| }\hline
Parameter & Value\\ \hline \hline
 \texttt{NBATCH}& 128 \\\hline
 \texttt{NCRITIC}& 5\\\hline
\texttt{LDIM}& 10 \\\hline
Optimizer& Adam (\texttt{lr}=1e-4,$\beta_1=0.5,\beta_2=0.9$)  \\\hline 
\end{tabular}
\end{center}
\end{table}

\section{The \jeansvii  Solver}
\label{JEAnS_section}

In this section we present an overview of the \jeansvii algorithm we developed for accurate mass estimates in spherically symmetric self-gravitating systems. 

The \jeans \citep{2017MNRAS.470.2034D} algorithm 
is a numerical solver that estimates the mass content and the kinematic profile of spherically symmetric gravitating systems. It models  independent of anisotropy, $\beta(r)$, assumptions and  it requires parametric functional forms for the mass density profiles. The best mass model is selected with the use of model selection criteria \citep{2017MNRAS.470.2034D} (Akaike Information Criterion, \citealt{sugiura1978further,opac-b1100695}, hereafter AICc). 
The radial velocity dispersion profile, $\sigmar$, is represented as a ``free form'' B-spline function, $\sigmar(r) = \sum_i a^i B_i(r)$. The correct kinematic profile is inferred from the data. The solver uses information of brightness and line-of-sight velocity moments, $\sigmalos$, to estimate marginalized distributions of the mass model parameters as well as the coefficients, $a^i$,  that describe the radial dispersion profile, $\sigmar$. 

The algorithm consists of three distinct phases. 
In the first phase it evaluates the \emph{simplest} 
kinematic profile that gives a satisfactory representation\footnote{That is, the best B-spline basis, $B_i(x)$, according to the bias-variance trade-off
 \citep{hastie01statisticallearning}.} to the data, as well as the most probable mass model. This is achieved  with the use of evolutionary optimization and quadratic programming. In the second 
phase, \jeans evaluates the optimum smoothing parameters from ideal theoretical models. Finally, in phase three the algorithm performs MCMC inference, for the determination of marginalized distributions of the  model parameters. 

The new version \jeansvii  
is significantly modified compared to the previously-published version \citep{2017MNRAS.470.2034D}. 
In the first phase we again evaluate the optimum B-spline basis, as well as the statistically most favoured mass model. We introduce a new quadratic programming formalism - the Dynamic Moments Solver (hereafter DMS) -  for the numerical solution of  the system of coupled integrodifferential   Equations (\ref{eq_Jeans_trad},\ref{eq_sigmaLOS_trad}). In the latest version of the \jeans we expand both the radial, $\sigmar$, and tangential, $\sigmat$, profiles in a B-spline basis of order $k=4$ (degree = 3)\footnote{The lower the degree of the B-spline basis, the smaller the condition number of the system of equations.}, i.e. 
$\sigmar(r) =\sigmarcoeff^i B_i(r)$, $\sigmat(r) =\sigmatcoeff^i B_i(r)$.  
This allows us to treat the Jeans equation as a local, 
$r_i$, constraint in the quadratic optimization problem of estimating the velocity moments, $\sigmar,\sigmat$. In combination with the local support of B-spline functions, this translates to more equations for the unknown coefficients $\sigmarcoeff^i,\sigmatcoeff^i$ that further reduce the feasible solution space. In comparison with the old version of the \textsc{JEAnS}, by solving the Jeans equation (Eq. \ref{DLI_fmontyB_Jeans})  with respect to $\sigmat$ and substituting under the integral sign of the $\sigmalos$ definition (Eq. \ref{JEAnS_sigmaLOS_def}), we loose the local equations that $\sigmarcoeff^i$ and $\sigmatcoeff^i$ coefficients participate after the last datum. By keeping the Jeans equation as a constraint we can evaluate equations for $\sigmarcoeff^i$ and $\sigmatcoeff^i$ in all space $r\in [0, \rvir]$. This has a direct positive impact on the quality of the recovered anisotropy profile, $\beta(r)$.

In a similar fashion to the first version of the \textsc{JEAnS}, we do not invert the dynamical equations, thus we avoid the problem of having to integrate/differentiate noisy numerical functions. We also include additional global and local  constraints that guarantee that the kinematic profiles lead to physically acceptable solutions ($\sigmar, \sigmat \geq 0,\; \forall r \in [0,\rvir)$).  
The fitness function is modified in order to include information from the full line-of-sight kinematics. The optimum smoothing parameters are now evaluated directly from the data according to the best bias-variance tradeoff using a validation data set, $D_{\mathrm{val}}$. The model selection is performed using a hold out test data set, $D_{\mathrm{test}}$.
In phase two we perform MCMC inference for the unknown stellar and DM mass model parameters, $\theta_{\star}, \theta_{\bullet}$. In this phase, the kinematic profile is treated as a nuisance parameter. Finally in the third phase, we perform stochastic programming (SP) in order to determine confidence intervals for  the velocity dispersion profiles, $\sigmar, \sigmat, \sigmalos$. 

In more detail (Fig. \ref{DLI_nat_JEAnS_diag}), the distinct phases  of the  \jeansvii are the following: 
\begin{enumerate}
\item  An evolutionary optimization (EA) phase.   
In this phase we determine: a) the simplest (best) B-spline basis\footnote{Equivalently, the knots $\xi_i$ that define the simplest basis.} for the representation of the unknown radial,  $\sigmar$, and tangential, $\sigmat$ velocity dispersions, (b)  the best candidate mass models and, (c), the best smoothing\footnote{The description of each of the four smoothing parameters, $\theta_{\mathrm{smooth}}$, is given in section \ref{tgJm_Objective_func_sect}.} parameters, 
$\theta_{\mathrm{smooth}}=\{
\lambda_1, \beta_1, \lambda_2, \beta_2
\}$. For the evaluation of the smoothing parameters we use a validation data set, $D_{\mathrm{val}}$, created from random sampling from the LOS $\sigma_{\mathrm{los}}^i$ marginalized distributions (see Fig. \ref{DLI_PlumCuspTan_DValidation}). The optimum smoothing parameters are the ones that minimize the validation error for all random samples, $\sigma_{\mathrm{los}}^i$. We give more details of this process in the section where we describe the fitness function. For the model selection, we use a ``hold-out'' LOS moments test data set, $D_{\mathrm{test}}$ (Section \ref{thJm_data_prep}), and we perform model selection (Section \ref{JEAnS_model_selection}) based on the out-of-sample prediction error (\emph{generalization test error}). This approach gives more robust model selection (in comparison with predictive information criteria
 \citep{Gelman2014}), since it heavily penalizes models that do not generalize well on unseen data.  \textcolor{black}{It should be stressed however, that the efficiency of the model selection process depends crucially on the number of available data points.}
\item A Markov Chain Monte Carlo (MCMC) analysis, keeping the B-spline basis and the smoothing parameters fixed, for the best mass model.  In this scheme, the radial and tangential coefficients, $\sigmarcoeff^i,\sigmatcoeff^i$ are treated as nuisance parameters: they are estimated at each iteration from the DMS.  This phase produces marginalized distributions of the parameters of stellar, $\theta_{\star}$, and DM, $\theta_{\bullet}$, mass densities, $\{\dtracer, \dDM \}$.  
\item A stochastic programming (SP) phase, where the $\theta_{\star},\theta_{\bullet}$ parameters are used iteratively in the DMS. This produces marginalized distributions for the radial and tangential coefficients,
 $\sigmarcoeff^i,\sigmatcoeff^i$, \emph{subject to local and global dynamical constraints}. This last phase gives the required uncertainty  of LOS and radial and tangential velocity dispersions. 
\end{enumerate}

\subsection{Mass models}
For our modelling purposes we used the following candidate mass models: 
\begin{equation}
\rho(r) =
\begin{cases}
 \dfrac{\rho_0}{ [1+(r/r_{\mathrm{s}})^2]^{5/2}}
&\mbox{Plummer}\\[10pt]
\dfrac{\rho_{0 }}{(1+r/r_{\mathrm{s}}) [1+(r/r_{\mathrm{s}})^2]} & \mbox{Burkert}\\[10pt]
   \dfrac{r_{\mathrm{s}}^3\rho_{0}}{r (r^2+r_{\mathrm{s}}^2)^{2}  } &\mbox{NFW}\\[5pt]
  \mathrm{Eq} \; \eqref{GAIA_mass_profile} & \mbox{generalized Hernquist}
  \end{cases}
\end{equation}
We model each dataset with two different mass model assumptions, the correct one and an incorrect one. Our goal is to demonstrate that given sufficient data it is possible, in principle,  to statistically infer the most probable model using model selection criteria.

\subsection{Dynamic Moments Solver (DMS)}
\label{DLI_DMS_sect}

In this section we describe the mathematical representation of the problem, i.e. the dynamic equations that enable us to recover the radial and tangential velocity moments, from knowledge of the LOS velocity dispersion, $\sigmalos$, the tracer, $\dtracer$, and the DM, $\dDM$, mass densities.  
The DMS  solves the system of coupled integrodifferential equations  
(Eq. \ref{eq_Jeans_trad},\ref{eq_sigmaLOS_trad})  by discretizing the solution space using B-splines. This is achieved by expanding the unknown radial, $\sigmar$, and tangential, $\sigmat$, velocity moments in a B-spline basis\footnote{We use Einstein summation convention, where double repeated indices indicate summation. E.g. $\sigmar (r) = a^iB_i(r) \equiv \sum_{i=1}^{n_{\mathrm{basis}}} a^iB_i $.} 
\begin{align}
\label{tgJm_sigmar_expansion}
\sigmar(r) &=\sigmarcoeff^i B_i(r)\\
\label{tgJm_sigmat_expansion}
\sigmat(r) &=\sigmatcoeff^i B_i(r)
\end{align}
The DMS takes as input the knots, $\xi_i$, the stellar parameters, $\theta_{\star}$, the DM parameters, $\theta_{\bullet}$ and the smoothing penalty variables, $\theta_{\mathrm{smooth}}=\{
\lambda_1, \beta_1, \lambda_2, \beta_2
\}$ and gives as output the coefficients  $\sigmarcoeff^i,\sigmatcoeff^i$ that fully describe the radial and tangential velocity moments. Using the approximation Eqs. \eqref{tgJm_sigmar_expansion} and \eqref{tgJm_sigmat_expansion}, the task is transformed to a convex optimization problem (quadratic programming). The software library we use in \jeansvii for the quadratic optimization is \href{http://www.ibm.com/software/products/ibmilogcpleoptistud/}{\textsc{IBM's CPLEX}}\footnote{Free academic license.}.

For clarity in notation, it is convenient  to represent the DMS as a function: 
\[
\mathrm{DMS}(\theta|D_{\mathrm{train}}) \to (\sigmarcoeff^i,\sigmatcoeff^i), 
\]
where $\theta \equiv \{\knots, \theta_{\star},\theta_{\bullet}, \theta_{\mathrm{smooth}} \}$ are the parameters that define the B-spline basis, the tracer and DM profiles, as well as the smoothing penalty regularization. 
The goal of the DMS is  to minimize the training error of the LOS velocity dispersion: 
\begin{equation}\label{DLI_training_error_kin}
\chi^2_{\mathrm{train}} =
\sum_i^{N_{\mathrm{bins}}}
\left(
\frac{
\sigmalos(R_i) - \sigma_{\mathrm{los}}^i}{\delta \sigma_{\mathrm{los}}^i}
\right)^2 
\end{equation}
subject to various local and global dynamic equations (constraints). We separate these  constraints into local, boundary and global  constraints.
In addition we will impose some regularization conditions (smoothing) in the minimization process, in order to reduce the condition number of the linear system and avoid oscillatory solutions. We formally define the objective function of the DMS in Section \ref{tgJm_Objective_func_sect}. 
In Table \ref{DIL_Summary_of_QP} we summarize the system of equations and the objective function that fully describe the DMS. We proceed by stating exactly the mathematical equations we use in the t-\textsc{JEAnS}.

The LOS velocity dispersion under the B-spline approximation of the velocity moments is given by: 
\begin{equation} \label{JEAnS_sigmaLOS_def}
\sigmalos  = \frac{1}{\Sigma_{\star}(R)} \biggl(
\int_{R}^{\rvir} \rho_{\star} K_1 \sigmar \diff r + 
\int_{R}^{\rvir} \rho_{\star} K_2 \sigmat \diff r \biggr)
\end{equation} 
where
\begin{equation*}
\Sigma_{\star}(R) = \int_R^{\rvir} \rho_{\star}  K_3 \diff x
\end{equation*}
is the projected tracer mass density 
and
\begin{align*}
K_1(r,R) &= \frac{2(r^2-R^2)}{r\sqrt{r^2-R^2}}\\
K_2(r,R) &= \frac{R^2}{r\sqrt{r^2-R^2}}\\
K_3(r,R) &= \frac{2 r}{\sqrt{r^2-R^2}}
\end{align*}
are kernel functions. 
Applying the B-spline approximation (Eq. \ref{tgJm_sigmar_expansion} and \ref{tgJm_sigmat_expansion}) and defining: 
\begin{align*}
 \IRlos_i(R) &= \frac{1}{\Sigma_{\star}}\int_R^{\rvir} \rho_{\star} K_1(r,R) B_i(r) \diff r\\
 \ITlos_i(R) &= \frac{1}{\Sigma_{\star}} \int_R^{\rvir} \rho_{\star} K_2(r,R) B_i(r) \diff r
\end{align*}
the linearized LOS velocity dispersion takes the form: 
\begin{equation}
\label{DLI_SFM_linearLOS}
\sigmalos(R)  =  \sigmarcoeff^i \IRlos_i(R) +  \sigmatcoeff^i \ITlos_i(R) 
\end{equation}
This is the model function that we compare with observables, subject to physical constraints. It is linear with respect to the unknown coefficients, $\sigmarcoeff^i , \sigmatcoeff^i $, something that simplifies the solution and allows for convex optimization. 

\subsection{Local constraints}
These constraints are termed local, because they are valid in the whole extent of the system, $r \in [0,\rvir]$. We  evaluate these at the positions of the Greville abscissae of the B-spline basis.

\subsubsection{Jeans constraints}
The spherically symmetric Jeans equation (SSJE) is: 
\begin{equation}\label{DLI_fmontyB_Jeans}
 -\rho_{\star}\frac{\diff \Phi}{\diff r}  = 
 \frac{\diff (\rho_{\star} \sigmar )}{\diff r} 
+\rho_{\star} \frac{(2 \sigmar - \sigmat)}{r}
\end{equation}
The linearized form of SSJE that results from the B-spline approximation is: 
\begin{equation}
\label{Jeans_FM_jeans}
-\rho_{\star} \frac{\diff \Phi}{\diff r} =
\left(
\frac{\diff (\rho_{\star} B_i)}{\diff r}  
+ \frac{2 \rho_{\star} B_i}{r} \right) \sigmarcoeff^i 
+ \frac{\rho_{\star}B_i}{r} \sigmatcoeff^i
\end{equation}

\subsubsection{Sign constraints}
We demand the velocity moments to be positive in all solution space:
\begin{align*}
\sigmar(r)  &\geq 0 \\
\sigmat(r)  &\geq 0. 
\end{align*}
In terms of the kinematic coefficients: 
\begin{align}
\sigmarcoeff^j B_j(r)  &\geq 0 \\
\sigmatcoeff^j B_j(r)  &\geq 0
\end{align}

\subsection{Boundary constraints} 
 These apply at the origin and at the virial radius of the system. 
\begin{align*}
 \sigmar(0)&=\sigmat(0)/2 \\
  \sigmar(\rvir)&=\sigmat(\rvir)=0 
\end{align*}
 The reasoning for the $\sigmar(0) =\sigmat(0)/2$ boundary condition is the following: we expect that all tangential motions at the limit $r \to 0$ become radial. That is, if we draw the tangent line to a circle of radius $r$, as the radius approaches zero, the tangent line approaches the origin $r=0$ of the coordinate system. In the limiting case where $r\to 0$ the tangent line passes from the origin (it is actually a degenerate case: all directions are equivalent). In this respect it is our understanding that in this limit the tangential and radial motions are indistinguishable. This is why we expect that their dispersions will be equal at $r \to 0$. With regards to the second boundary constraint, it is proven \citep{1992ApJ...391..531D} that for a self consistent system in virial equilibrium the radial and tangential velocity dispersions vanish in the limit of the virial radius.

\subsection{Global constraints}
In this category fall constraints of local functions integrated over all space.  
\subsubsection{Projected virial theorem}
\label{sec:Projected_virial_theorem}

The virial theorem states \citep{2008gady.book.....B,2013degn.book.....M} that if $K$ is the total kinetic energy of a system, and $\mathcal{W}$ its total potential energy, then for a system in dynamic equilibrium: 
\begin{equation}
\label{JEAnS_virial_original}
2 K + \mathcal{W} = 0
\end{equation}
For a sperically symmetric system, 
\begin{align*}
\mathcal{W} &=  4 \pi  \int_0^{\rvir}\rho_{\star}    \left( -\frac{\diff \Phi}{\diff r} \right) r^3 \diff r
\end{align*}
The total kinetic energy of a system, defined via the line-of-sight velocity dispersion is: 
\begin{equation*}
\Klos =  \frac{3}{2} \int_{R=0}^{\rvir}  \diff R\, 2 \pi  R \pdtracer(R)\; \sigmalos (R)
\end{equation*}
Substituting $\sigmalos (R)$  from Eq. \eqref{DLI_SFM_linearLOS}, we have
\begin{equation}
\Klos = \sigmarcoeff^i \KRlos_i + \sigmatcoeff^i \KTlos_i
\end{equation}
where
\begin{align*}
\KRlos_i  &= 3 \pi  \int_{R=0}^{\rvir} R \pdtracer(R)   \IRlos_i(R)\ {\rm d}R \\
 \KTlos_i  &=3 \pi    \int_{R=0}^{\rvir} R \pdtracer(R) \ITlos_i(R)\ {\rm d}R 
\end{align*}
Substituting in Eq. \eqref{JEAnS_virial_original} yields: 
\begin{equation}
\label{Jeans_FM_virial_final}
2 (K^{\mathrm{Rlos}}_i \sigmarcoeff^i + \KTlos_i \sigmatcoeff^i)+
 \mathcal{W}  = 0.
\end{equation}
This is an additional constraint on the $\sigmarcoeff^j,\sigmatcoeff^j$ coefficients.
From the perspective of linear/quadratic programming algorithmic structure,   
Equation \eqref{Jeans_FM_virial_final} is a hyperplane equation with respect to the unknown coefficients, $\sigmatcoeff^j$, $\sigmarcoeff^j$, that further reduces feasible solution space. 

The projected virial theorem is also a hard bound on the value of the total gravitational energy of the stellar and dark matter ($\star\bullet$) interaction. Furthermore, it is clear that since the $\Klos$ value is independent of the anisotropy profile (i.e. it is an observational fact), then it is impossible to have only the stellar component with some peculiar anisotropy profile to represent the observables. In other words, the total gravitational energy of the system is fixed from the total kinetic energy as this is  estimated from the LOS dispersion, $\sigmalos$.   That is, the constraint of virial equilibrium does not allow one to vary the anisotropy profile $\beta$ to fit any desired mass profile.

\subsection{Objective Function}
\label{tgJm_Objective_func_sect}

The  DMS objective function that relates observables, $\sigmalos$, with the model function (Eq. \ref{DLI_SFM_linearLOS}) is given by: 
\begin{multline}
\label{DLI_Object_Function}
\mathcal{F} = 
\sum_i
\left(
\frac{
\sigmalos(R_i) - \sigma_{\mathrm{los}}^i}{\delta \sigma_{\mathrm{los}}^i}
\right)^2  \\
+ \lambda_1 [ \beta_1 \sum_{i=1}^{n_{\mathrm{coeff}}-1}  (\Delta^1 \sigmarcoeff^i)^2 + (1-\beta_1) \sum_{i=1}^{n_{\mathrm{coeff}}-2}(\Delta^2 \sigmarcoeff^i)^2] 
\\ 
 + \lambda_2 [ \beta_2 \sum_{i=1}^{n_{\mathrm{coeff}}-1}  (\Delta^1 \sigmatcoeff^i)^2 + (1-\beta_2) \sum_{i=1}^{n_{\mathrm{coeff}}-2} (\Delta^2 \sigmatcoeff^i)^2] 
\end{multline}
where  the difference operators $\Delta^{1,2}$ are defined by: 
\begin{align*}
\Delta^1 c^i_{\mathrm{r,t}} &= c_{\mathrm{r,t}}^{i+1} - c_{\mathrm{r,t}}^i \\
\Delta^2 c_{\mathrm{r,t}}^i &= \Delta^1 (\Delta^1 c_{\mathrm{r,t}}^i) 
= c_{\mathrm{r,t}}^{i+2} - 2c_{\mathrm{r,t}}^{i+1} c_{\mathrm{r,t}}^{i} + c_{\mathrm{r,t}}^{i}.
\end{align*}
The coefficients, $\lambda_{1,2}$ regulate the ammount of smoothing penalty on each of the velocity dispersions. The coefficients $\beta_{1,2}$ regulate the relative contribution of the first and second derivative penalties for each velocity dispersion. 
This smoothing penalty is efficient and very fast to evaluate in comparison with previous efforts 
 \citep{2014MNRAS.443..610D,2017MNRAS.470.2034D}. It is the same penalty used in the  P-splines \citep{Eilers96flexiblesmoothing} formulation in statistical smoothing.

\begin{table*}
\caption{Summary of the  DMS solver $\left( \DMS (\theta|D) \to (\sigmarcoeff,\sigmatcoeff)\right)$  in the  \jeans modelling approach. \newline
{\bf Assumptions:} \textcolor{black}{(a) Spherical symmetry, (b) virial equilibrium, (c) parametric form for the stellar and DM mass profiles.} \newline
{\bf Input:} $\theta \equiv \{ \theta_{\star},\theta_{\bullet}, \xi,\smoothLambda_{1,2},\smoothBeta_{1,2} \}$ and training data, $D=\{R_i, {\sigma}_{\mathrm{los}}^i , \delta {\sigma}_{\mathrm{los}}^i \}$. \newline
{\bf Output:} $\sigmat = \sigmarcoeff^i B_i,\sigmar = \sigmatcoeff^i B_i$. }
\label{DIL_Summary_of_QP}
\begin{center}
\begin{tabular}{ | l |l| c |   }
\hline \hline
& &  Mathematical formula \\
\hline\hline
Objective function   & &
 $\min \mathcal{F} = 
\sum_i(\sigmalos(R_i) 
-\boldsymbol{ \sigma}_{\mathrm{los}}^i)^2
+ \lambda_1 [ \sum_{i=1}^{n_{\mathrm{coeff}}-1} \beta_1 (\Delta^1 \sigmarcoeff^i)^2 + (1-\beta_1) (\Delta^2 \sigmarcoeff^i)^2] $ \\  
& &  $ + \lambda_2 [\sum_{i=1}^{n_{\mathrm{coeff}}-2} \beta_2 (\Delta^1 \sigmatcoeff^i)^2 + (1-\beta_2) (\Delta^2 \sigmatcoeff^i)^2]$\\[7pt]
Model function & & $\sigmalos(R) = \sigmarcoeff^i \IRlos_i(R) + \sigmatcoeff^i \ITlos_i(R) $\\[5pt] \hline\hline
Local constraints & & \\\hline \hline
& Jeans: & $-\rho_{\star} \dfrac{\diff \Phi}{\diff r}  =
\left(
\dfrac{d (\rho_{\star} B_i)}{dx}  
+ \dfrac{2 \rho_{\star} B_i}{x} \right) \sigmarcoeff^i 
+ \dfrac{\rho_{\star}B_i}{x} \sigmatcoeff^i$\\[5pt] \hline
& sign: & $\sigmarcoeff^j B_j(x)  \geq 0 , \sigmatcoeff^j B_j(x)  \geq 0$ \\[5pt] \hline
& boundary: & $ \sigmar(0)=\sigmat(0)/2$\\[5pt]   
& & $\sigmar(\rvir)=\sigmat(\rvir)=0 $\\[5pt]\hline \hline
Global constraints  & & \\ \hline\hline 
& projected virial: & $2 (\KRlos_i \sigmarcoeff^i + K^{\mathrm{Tlos}}_i \sigmatcoeff^i)+ W = 0$\\ \hline\hline
\end{tabular}
\end{center}
\end{table*}

\subsection{Fitness function}
The EA phase of the \jeansvii  solver evaluates the
simplest B-spline basis that best represents the observables. This is a nested optimization: the EA parameters consist of the stellar, $\theta_{\star}$, the DM, $\theta_{\bullet}$ and the smoothing penalty variables, $\theta_{\mathrm{smooth}}=\{
\lambda_1, \beta_1, \lambda_2, \beta_2
\}$. Once these parameters, $\theta = \{\theta_{\star}, \theta_{\bullet}, \theta_{\mathrm{smooth}} \}$,  are proposed, then the problem is a quadratic programming optimization problem, with respect to the $\sigmarcoeff^i,\sigmatcoeff^i$ unknown constants. The optimal variables, 
$\hat{\sigmarcoeff}^i,\hat{\sigmatcoeff}^i$ for the proposed $\theta$ parameters are evaluated with the DMS. The evaluation of the model though, takes into account information from both the DMS and the full kinematics. For the full kinematics, we use definitions \citep{2013MNRAS.429.3079M} based on assumptions of a Gaussian distribution function for the  velocities (in 3D space), truncated at the escape velocity of the system.

The fitness function is defined with the usage of model selection criteria ($\BIC$, $\AICc$) and the following penalty functions: 
\begin{align*}
\AICc &= 2 \sum_i
\frac{1}{2}
\left(
\frac{
\bestsigmalos(R_i) - \sigma_{\mathrm{los}}^i}{\delta \sigma_{\mathrm{los}}^i}
\right)^2 + 2 n + \frac{2n (n+1)}{N_{\mathrm{data}} - n - 1}\\
\chi_{\mathrm{smooth}} &= \frac{1}{ N_{\mathrm{sample} } } \sum_{j=1}^{ N_{\mathrm{sample} }}
 \sum_{i=1}^{N_{\mathrm{bins}}}
 \left(
 \frac{
\bestsigmalos(R_i) - \sigma_{\mathrm{los}}^{ij}
}{ \delta \sigma_{\mathrm{los}}^i}
\right)^2 \\
\chi_{\mathrm{virial}} &=
\left| \frac{2 \Klos}{W} - \frac{W}{2\Klos} \right|  \\
\mathrm{BIC} &=
-2\sum_{i=1}^{N_{\mathrm{batch}}} 
 \log q(R_i,v_{\los}^i)) + n \log (N_{\mathrm{batch}})  
\end{align*}
where   
\begin{align}
 q(R,v_{\mathrm{los}}) &= \frac{2 \pi R}{
 M^{\star}_{\mathrm{tot}}} g(R,v_{\los}) 
 \\
 g(R,v_{\los}) &= \int_R^{\rvir}
 \frac{r \dtracer(r)}{\sqrt{r^2-R^2}} h(v_{\los}|R,r) \d r\\
 h(v_{\los}|R,r) &= \frac{\exp[-\frac{v_{\los}^2}{2 \sigma_z^2(R,r)}]}{\sqrt{2 \pi \sigma_z^2(R,r)} 
 \erf\{ 
v_{\mathrm{esc}}(R)/\sqrt{2 \sigma_z^2(R,r)}\} }
 \\
 \sigma_z^2(R,r) &= \sigmar (1-(R/r)^2)+\sigmat (R/r)^2/2
\end{align}
are the  full kinematics definitions from the \mampost\ \citep{2013MNRAS.429.3079M} algorithm.  The projected virial theorem  is satisfied by the quadratic programming solver, within some numerical tolerance. We found that we got slightly faster convergence  by also penalizing this explicitly in the EA solver, with  the $\chi_{\mathrm{virial}}$ term.

The values $\bestsigmalos(R_i)$ are the solutions from the Dynamical Modelling Solver (DMS) for the given input parameters $\theta = \{ \boldsymbol{\xi}, \theta_{\star}, \theta_{\bullet}, \lambda_{1,2}, \beta_{1,2}\}$.
 The data values, $\sigma^i_{\mathrm{los}}$ and  
$\sigma_{\mathrm{los}}^{ij}$ are produced from a binning scheme as described in Section  \ref{thJm_data_prep}.  We remind the reader that the values  $\sigma_{\mathrm{los}}^{ij} \in D_{\mathrm{val}}$  are $j$ sampled values for each bin $i$. They are used as a validation set for determining the smoothing parameters, $\theta_{\mathrm{smooth}}=\{
\lambda_1, \beta_1, \lambda_2, \beta_2
\}$. 
$N_{\mathrm{batch}}$ is the size of a  random sample (without replacement) of full kinematics data of  stars from the population. We use  $N_{\mathrm{batch}}=1000$: this provides a good approximation to the full kinematics likelihood and allows for faster convergence.

\begin{figure}
\centering
\includegraphics[width=\columnwidth]{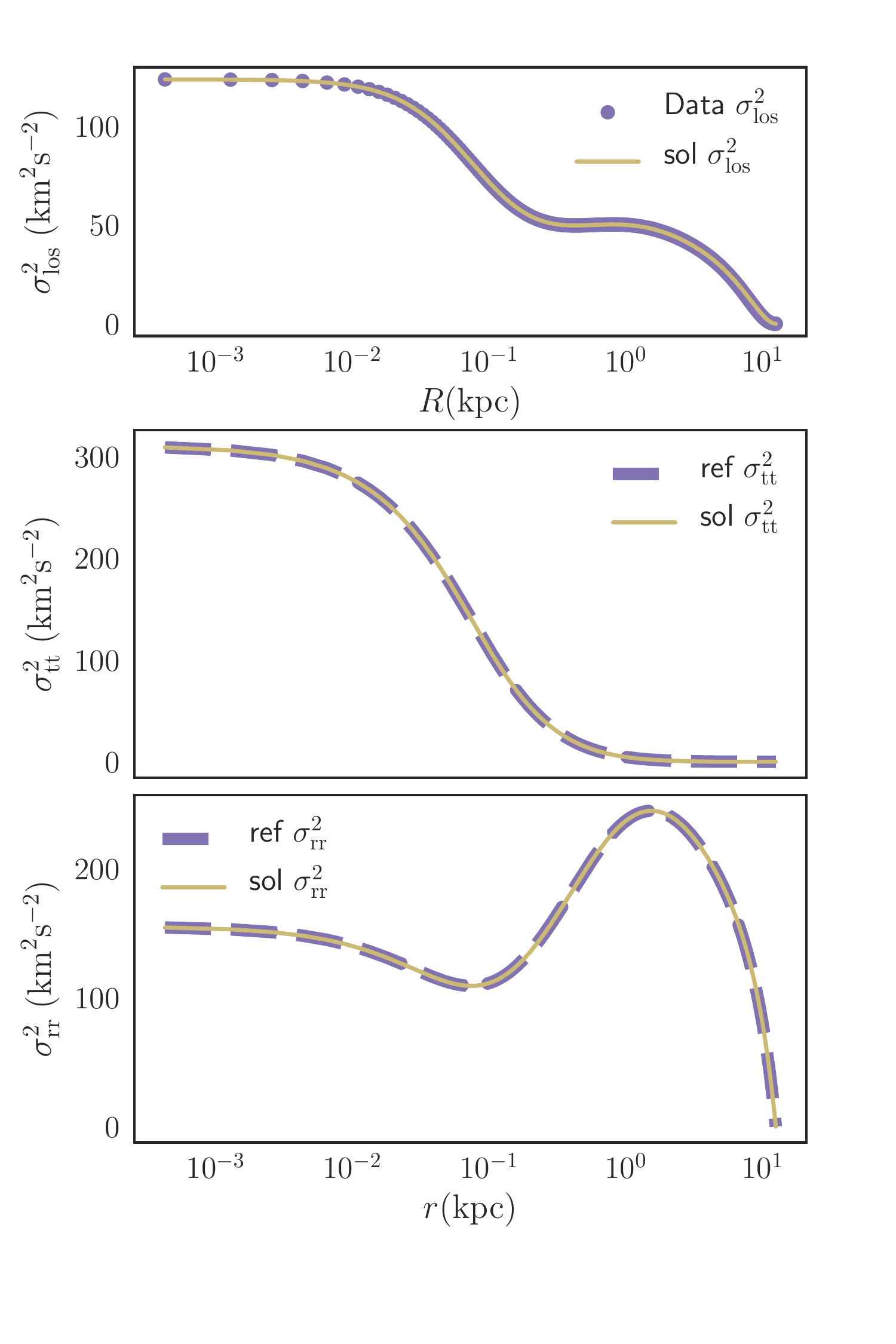}
\caption{  Exact numerical solution of the system of Jeans equations for the PlumCuspOM model using the Dynamic Moments Solver (DMS, see Table \ref{DIL_Summary_of_QP}). In order to obtain this solution we  assumed perfect knowledge of the $\sigmalos,\dtracer$ and $\dDM$ profiles.}
\label{DLI_PlumCuspOM_POC}
\end{figure}

\begin{figure}
\centering
\includegraphics[width=\columnwidth]{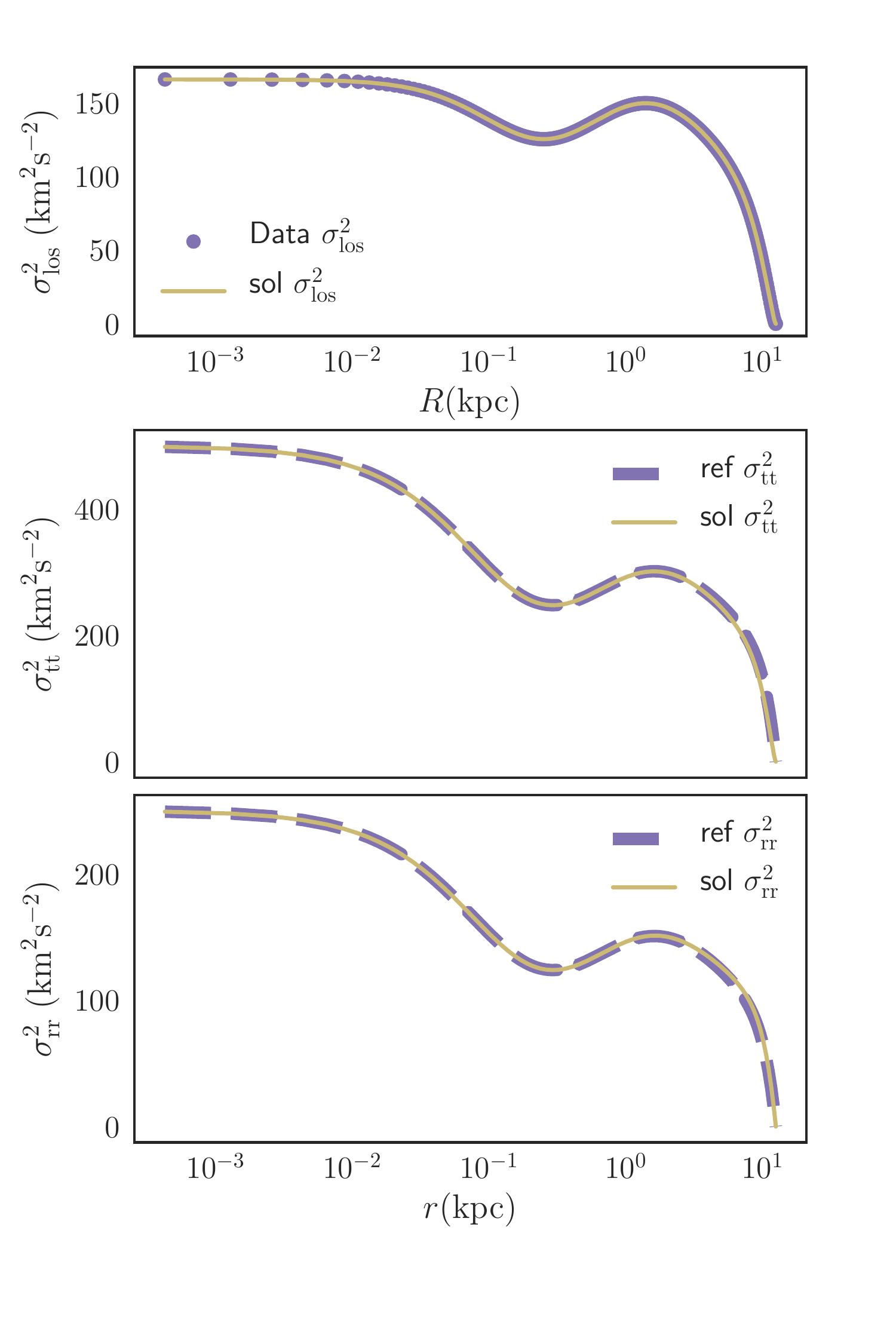}
\caption{  As Fig. \ref{DLI_PlumCuspOM_POC} for the PlumCuspIso reference profile.}
\label{DLI_PlumCuspIso_POC}
\end{figure}

The fitness function, treated as a maximization problem, is the product of four components, namely:
\begin{align*}
f_{\mathrm{DMS}} &= \frac{1}{1+\AICc} \\
f_{\mathrm{full\, kin}} &= \frac{1}{1+\BIC} \\
f_{\mathrm{smooth}} &= \frac{1}{1+\chi_{\mathrm{smooth}}}\\
f_{\mathrm{vir}} &= \frac{1}{1+\chi_{\mathrm{vir}}}.
\end{align*}
Then, the fitness function is: 
\begin{equation}
\label{DLI_tgJm_tot_fitness}
f_{\mathrm{tot}}(\theta) = f_{\mathrm{DMS}}\; 
f_{\mathrm{full\, kin}}\;  f_{\mathrm{smooth}} \; f_{\mathrm{vir}}
\end{equation}

\subsection{Model selection }
\label{JEAnS_model_selection}

Model selection takes place in two distinct processes inside the t-\textsc{JEAnS}. Once we select a set of tracer and DM mass densities, we use the EA in order to find the simplest B-Spline basis for the radial and tangential velocity dispersions. This task is a hierarchical model selection problem (where the various competing models are the ones that have different number and locations of knots, but the same mass density parametric form). For this task, AICc or BIC, based on the training  error measure (likelihood) prove to be good choices. 

However, when one needs to compare competing mass models that were trained in distinct EA phases,   it is best to use out-of-sample data and test how well the model generalizes on unseen (during training) data (Section \ref{JEAnS_data}). 
Once the EA phase is complete, for competing mass models, we evaluate the best model using the hold-out LOS moments test set in a cross-validation manner. The average error on unseen moments test data that we use is: 
\begin{equation}
\label{tJEAnS_test_error_model_selection}
\chi^2_{\mathrm{test}} = \frac{1}{2 N_{\mathrm{test}}} \sum_{j=1}^{N_{\mathrm{test}}} \sum_{i=1}^{N_{\mathrm{bin}}}
\left(
\frac{
\bestsigmalos(R_i) - \tilde{\sigma}_{\mathrm{los}}^{ij}}{\delta \sigma_{\mathrm{los}}^i}
\right)^2
\end{equation}
where $\tilde{\sigma}_{\mathrm{los}}^{ij}$ are out of sample test data, created for each bin $i$ by random sampling from the marginalized distribution of the data preprocessing MCMC chains. The model with the smallest test error is selected as the best candidate. In our experiments this method has proven to be more robust than predictive training error methods (e.g. AICc), which can have bias from overfitting \citep{Gelman2014}.

\begin{figure*}
\centering
\includegraphics[width=\textwidth]{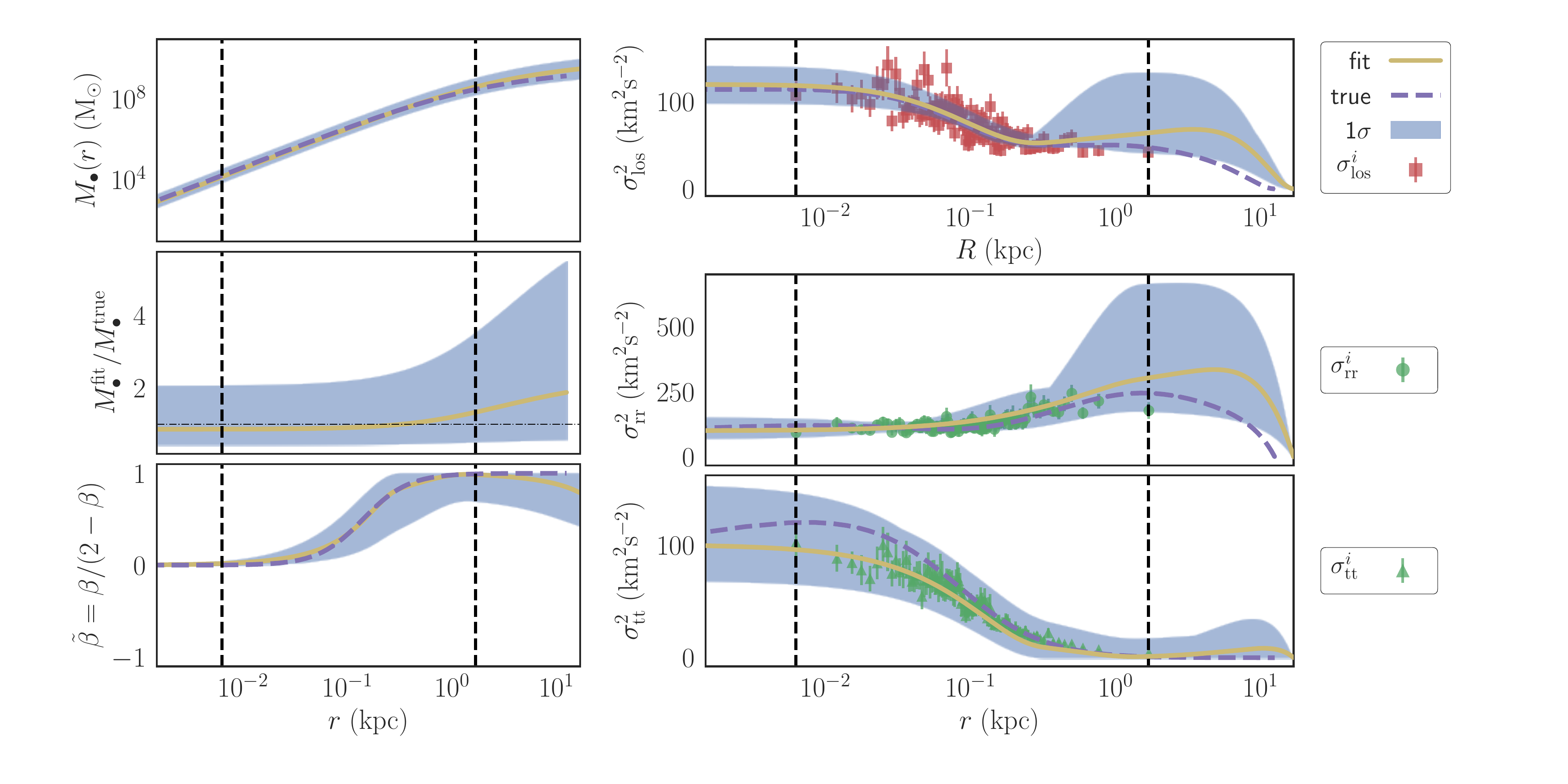}
\caption{ Fit and true reference profiles for the GAIA Challenge dataset PlumCuspOM, for 10k targets. In the left panels we plot (top) the DM mass, (middle) the ratio of the estimated  mass over the true mass and (bottom) the anisotropy profile. In the right panels we plot (top) the fit to the LOS observables, (middle and bottom) the recovered radial and tangential profiles. Overplotted are the true $\sigma_{\mathrm{rr}}^i,\sigma_{\mathrm{tt}}^i$ dispersions, as estimated from the data; these were not used in the fitting process. The blue region corresponds to 1$\sigma$  uncertainty for all of the quantities.}
\label{DLI_nat_PlumCuspOM_10k}
\end{figure*}

\subsection{Likelihood function}
In Phase II of the \jeansvii we perform an MCMC exploration using the following likelihood \citep{2013MNRAS.429.3079M,2013MNRAS.428.3648I,2017MNRAS.470.2034D,2014MNRAS.443..598D,2014MNRAS.443..610D}: 
\begin{align}
\notag
\mathcal{L} =
& \textcolor{black}{\left[ \prod_{j=1}^{N^{\star}_{\mathrm{bin}}}  
\frac{\exp\{ -
\frac{\left(\Sigma_{\star}(R_j) - \Sigma^j\right)^2}{2 (\delta\Sigma_{\star}^j)^2 }
\}}{\sqrt{2 \pi (\delta\Sigma_{\star}^j)^2 } }
\right] \times} \\
& \left[ \prod_{i=1}^{N_{\mathrm{bin}}}  
\frac{\exp \left\{  
-
\frac{\left(
\bestsigmalos(R_i) -\tilde{\sigma}_{\mathrm{los}}^i \right)^2}{\textcolor{black}{2}\delta (\sigma_{\mathrm{los}}^i)^2}
  \right\}}{\sqrt{2\pi \delta (\sigma_{\mathrm{los}}^i)^2}}\right] \times \\
&\times \lambda_1 e^{-\lambda_1 W_1} \lambda_2 e^{-\lambda_2 W_2} 
\left(\prod_{j=1}^{N_{\mathrm{batch}}} q(R_j,v_{\mathrm{los}}^j)
\right)
\end{align}
where {$\Sigma_{\star}(R_j)$ is the projected tracer density at location $R_j$, $ \Sigma^j_{\star}$ and $\delta \Sigma^j_{\star}$ the observed projected mass density and its uncertainty},  
$\tilde{\sigma}_{\mathrm{los}}^i$ is a random sampled value (at each iteration, we use values from $D_{\mathrm{val}}$) from the $i$th MCMC binned histograms and  $W_1, W_2$ are given by:
\begin{align}
W_1 &=  \beta_1 \sum_{i=1}^{n_{\mathrm{coeff}}-1}  (\Delta^1 \sigmarcoeff^i)^2 + (1-\beta_1)\sum_{i=1}^{n_{\mathrm{coeff}}-2} (\Delta^2 \sigmarcoeff^i)^2 \\
W_2 &=  \sum_{i=1}^{n_{\mathrm{coeff}}-1} \beta_1 (\Delta^1 \sigmatcoeff^i)^2 + (1-\beta_1) \sum_{i=1}^{n_{\mathrm{coeff}}-2}(\Delta^2 \sigmatcoeff^i)^2 
\end{align}
The B-spline knots, and the coefficients $\lambda_{1,2}, \beta_{1,2}$ are kept fixed to the values of the best EA solution. The full kinematics likelihood is calculated on each iteration on a random sample (without replacement) of $N_{\mathrm{batch}}=1000$ stars. This is sufficient for the algorithm to converge in an excellent trade-off between computational efficiency and parameter constraints. We use random samples, $\tilde{\sigma}_{\mathrm{los}}^i$, as data in each MCMC iteration in order to avoid overoptimistic constraints for the marginalized distributions of parameters. In this way we incorporate the uncertainty of the binned LOS dispersion values in the marginalized distributions of the $\theta_{\star},\theta_{\bullet}$ parameters.

\begin{figure*}
\centering
\includegraphics[width=\textwidth]{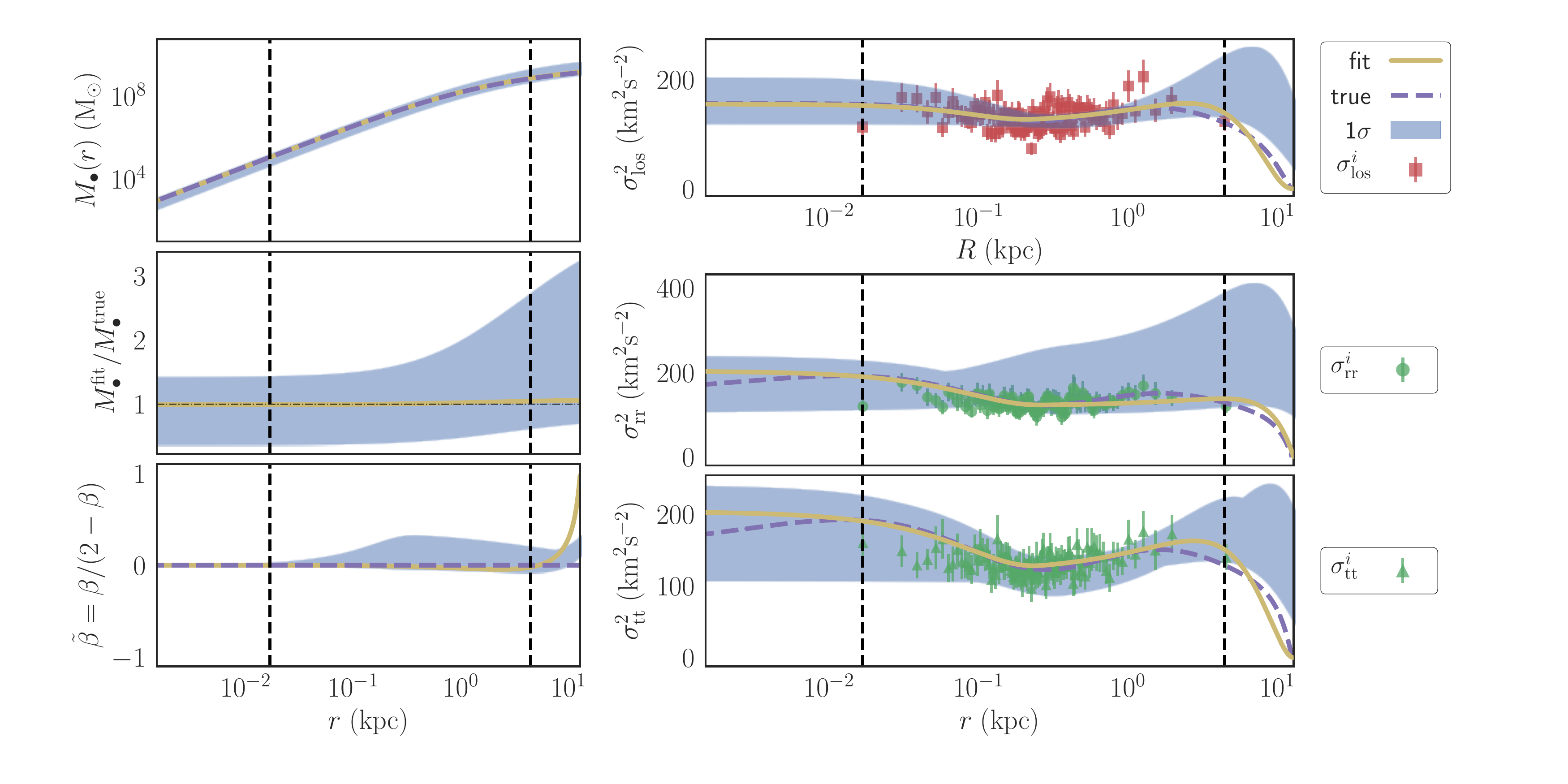}
\caption{ Same as Fig \ref{DLI_nat_PlumCuspOM_10k} for the  PlumCuspIso 10k dataset. }
\label{DLI_nat_PlumCuspIso_10k}
\end{figure*}

\subsection{Stochastic Programming}
\label{JEAnS_SP_section}
Once we have clean (after burn-in phase) MCMC chains of the 
$\theta_{\star},\theta_{\bullet}$ parameters, we estimate distributions of $\sigmarcoeff, \sigmatcoeff$ by applying the DMS solver iteratively to each pair of MCMC values $\theta_{\star}^j,\theta_{\bullet}^j$. Here, the index $j$ indicates the $j$th MCMC chain. For this computation we keep the smoothing penalty parameters, as well as the B-spline basis, fixed to the best EA values. In functional form: 
\begin{equation}
\mathrm{PDF}(\sigmarcoeff^i, \sigmatcoeff^i) = 
\mathrm{DMS}\left(\theta_{\star}^j,\theta_{\bullet}^j \sim \mathrm{PDF}(\theta_{\star},\theta_{\bullet})| D \right)
\end{equation}
where the symbol $\theta_{\star}^j,\theta_{\bullet}^j \sim \mathrm{PDF}(\theta_{\star},\theta_{\bullet})$ denotes that $\theta_{\star}^j,\theta_{\bullet}^j$ are sampled at random from their marginalized distribution $\mathrm{PDF}(\theta_{\star},\theta_{\bullet})$ (estimated from the MCMC chains). 
Finally, from the marginalized distributions of $\sigmarcoeff, \sigmatcoeff,\theta_{\star},\theta_{\bullet}$, we can estimate 1$\sigma$ uncertainty intervals for the velocity moments and the various mass model functions.

 \section{Results} 
 \label{section_results}
In this section we summarize our findings for both the exact solution of the Jeans system of equations and the  statistical fitting of the \GAIA dataset. 

\subsection{Exact solutions}
\label{section_results_exact}

In Figures \ref{DLI_PlumCuspOM_POC} and \ref{DLI_PlumCuspIso_POC} we plot the exact solutions of the system of the Jeans equations (Table \ref{DIL_Summary_of_QP}) using the DMS solver, for the case of the PlumCuspOM and PlumCuspIso reference profiles. Our aim here is to provide  numerical ``proof of concept'' examples of Theorem \ref{JEAnS_theorem}. That is,  by assuming full knowledge of the LOS velocity dispersion profile, the tracer $\dtracer$ and $\dDM$ mass densities, we recover a unique kinematic profile as this is described by the second order radial, $\sigmar$, and tangential, $\sigmat$, velocity moments. In this approach we are not using smoothing penalty coefficients ($\smoothLambda_{1,2}=0$ in the objective function Eq \ref{DLI_Object_Function}) since we have a wealth of data points. For the exact solution we use a large number of B-Spline basis, $\dim \{ B_i(r) \} \sim 150$. For each of the two figures, from top to bottom panels: data $\sigmalos$ and recovered solution, reference and recovered tangential velocity dispersion ($\sigmat$) and reference and recovered radial velocity dispersion profile ($\sigmar$).

\subsection{Statistical fitting}

\begin{figure*}
\centering
\includegraphics[width=\textwidth]{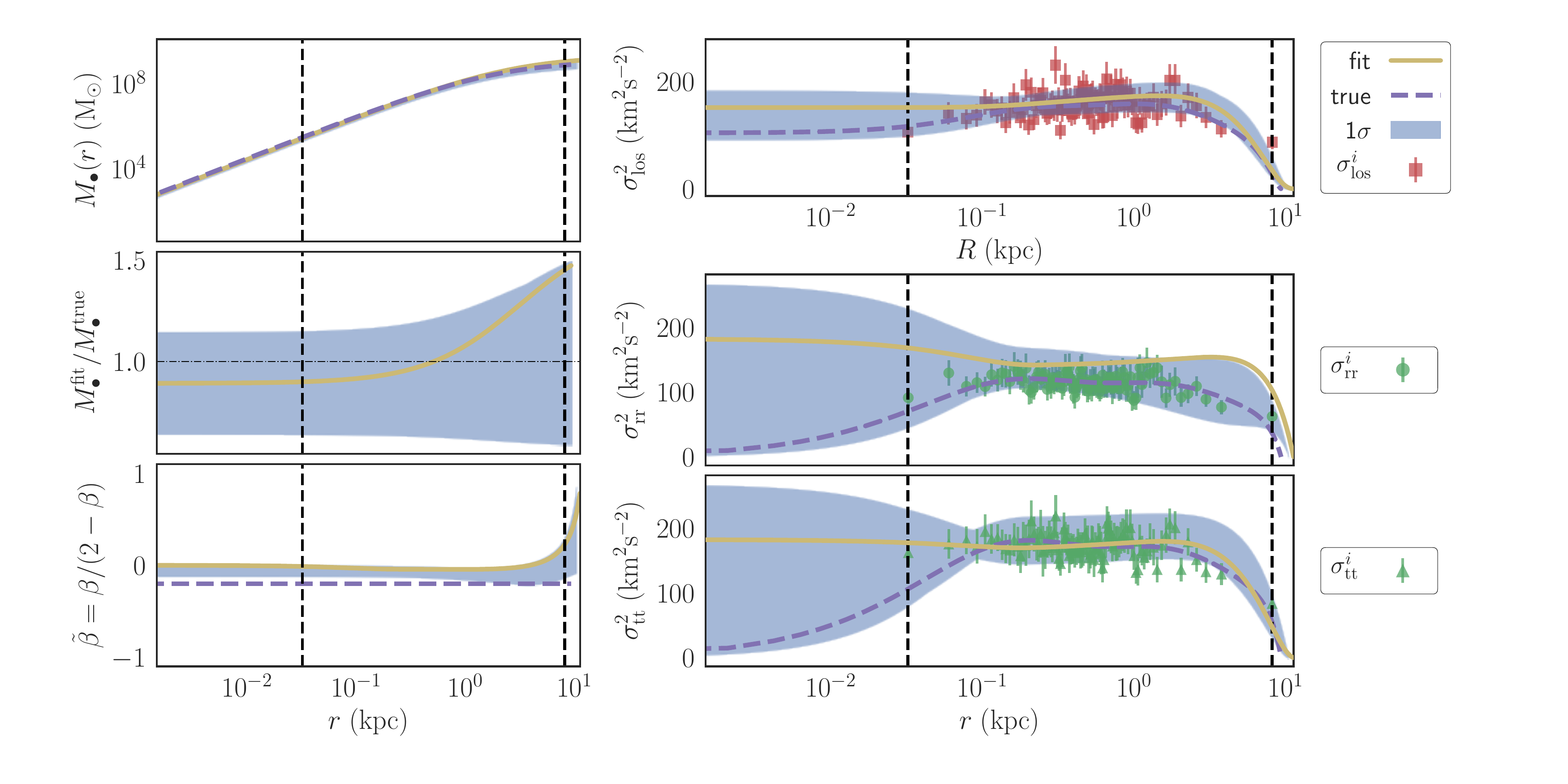}
\caption{ Same as Fig \ref{DLI_nat_PlumCuspOM_10k} for the  PlumCuspTan 10k dataset. }
\label{DLI_nat_PlumCuspTan_10k}
\end{figure*}

\begin{figure*}[ht]
\centering
\includegraphics[width=\textwidth]{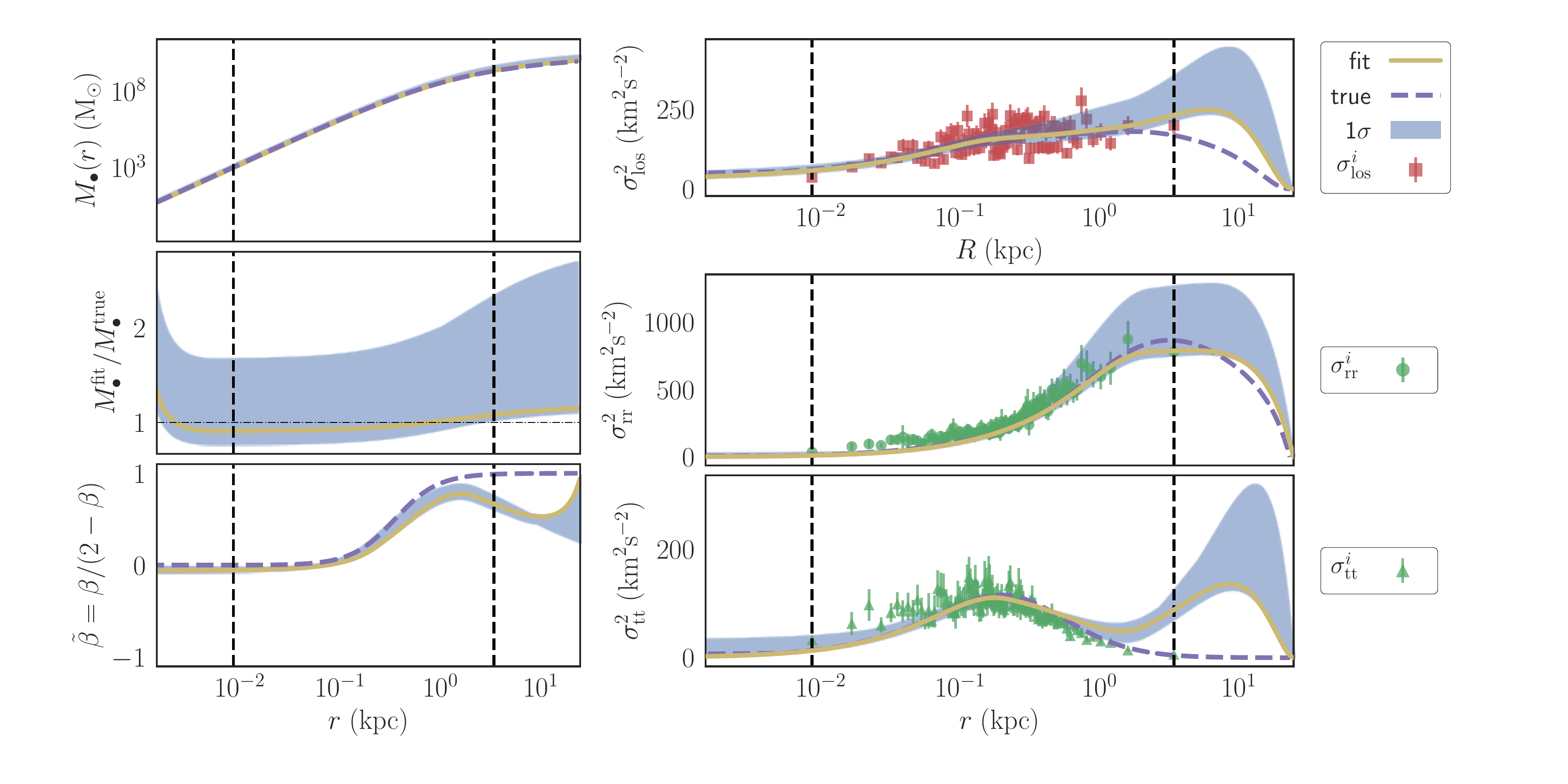}
\caption{ Same as Fig \ref{DLI_nat_PlumCuspOM_10k} for the  NonPluCoreOM 10k dataset. }
\label{DLI_nat_NonPlumCoreOM_10k}
\end{figure*}

\begin{figure*}[ht]
\centering
\includegraphics[width=\textwidth]{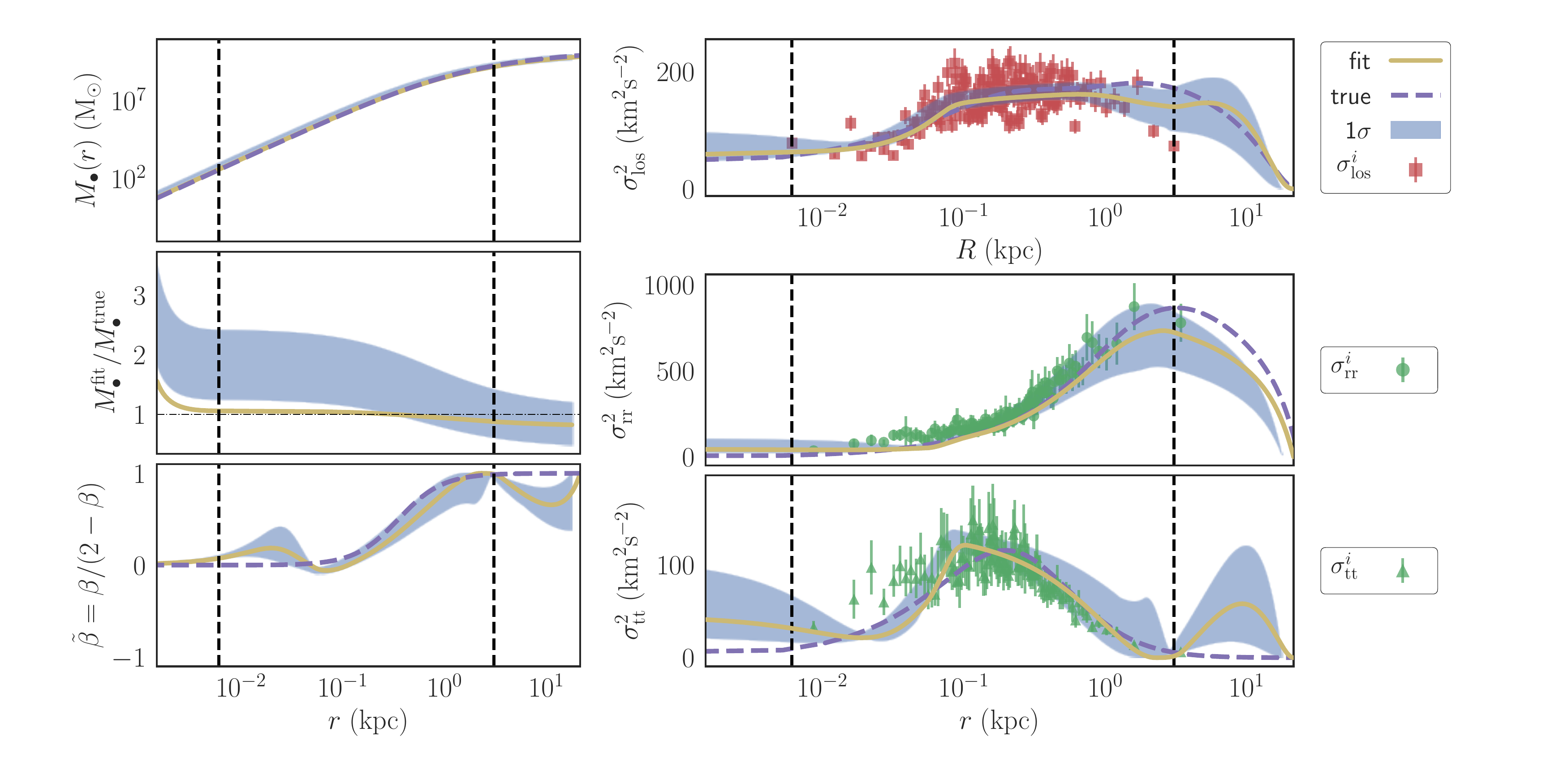}
\caption{ Same as Fig \ref{DLI_nat_PlumCuspOM_10k} for the  NonPluCoreOM 1k dataset. In the top right panel the $\sigmalos$ data are from the GAN synthetic data generator. In the two bottom right panels the $\sigmar$ and $\sigmat$ moments are from the true NonPlumCoreOM 10k dataset (we do not have 3D motions for the GAN generated data). }
\label{DLI_nat_NonPlumCoreOM_1kGAN}
\end{figure*}

 \subsubsection{10k datasets}
 
  Our results are summarized in Figures  \ref{DLI_nat_PlumCuspOM_10k}, \ref{DLI_nat_PlumCuspIso_10k}, \ref{DLI_nat_PlumCuspTan_10k} and \ref{DLI_nat_NonPlumCoreOM_10k} and Table \ref{JEAnS_test_models_2}. We fully recover the mass content and the anisotropy profiles in a representative sample of synthetic data sets from the \GAIA\footnote{http://astrowiki.ph.surrey.ac.uk/dokuwiki/doku.php} suite of mock simulations. 
In Fig. \ref{DLI_nat_PlumCuspOM_10k} we plot the best fitting model, as well as the 1$\sigma$ uncertainty interval for a data set with Plummer like tracer profile, a Cuspy DM halo and Ossipkov-Merritt \citep{1979SvAL....5...42O,1985AJ.....90.1027M} velocity anisotropy profile (PlumCuspOM), for 10k stars. In all panels, 
the vertical dashed lines designate the values of the first and last datum.  The reliable region for making predictions is within these lines. Everything outside this region is extrapolation and cannot be trusted. Left panels, from top to bottom:  estimated DM mass,   the ratio of the fitted to the true DM mass and  the normalized \citep{2017arXiv170104833R} velocity-anisotropy. Right panels, from top to bottom: LOS velocity dispersion fit and the data we used. The radial (middle), $\sigmar$,  and tangential (bottom), $\sigmat$, velocity moments. The data in the middle and bottom panels were not used in the fitting process. They are produced from the true 3D kinematic information and are shown for comparison with the fitted models. 

Figures \ref{DLI_nat_PlumCuspIso_10k}, \ref{DLI_nat_PlumCuspTan_10k} and \ref{DLI_nat_NonPlumCoreOM_10k} are as Fig. \ref{DLI_nat_PlumCuspOM_10k} for  the \GAIA datasets with: a Plummer like tracer profile with cuspy dark matter halo and isotropic velocity anisotropy  (PlumCuspIso), a Plummer like tracer profile with cuspy DM halo and tangential velocity anisotropy (PlumCuspTan) and a cuspy like (non-Plummer) tracer profile with a cored halo and Ossipkov-Merritt
 \citep{1979SvAL....5...42O,1985AJ.....90.1027M} velocity anisotropy profile (NonPlumCoreOM) datasets. In all four cases, our algorithm selects the correct model and reconstructs robustly the mass content and kinematic profile of the underlying stellar distributions from LOS data only.

 In Table \ref{JEAnS_test_models_2} we report the results of the mass model selection during Phase I of the t-\textsc{JEAnS}. The model selection is performed using the average test error on unseen (during training) data (Eq \ref{tJEAnS_test_error_model_selection}) as it has proven to be a more robust discriminator (in comparison with AICc or BIC). We perform model selection after Phase I, in order to  reduce computation time. In general, better discrimination results between competing models can be achieved by performing the MCMC process (Phase II) for both competing models and then evaluating the test error, $\chi^2_{\mathrm{test}}$  (Eq \ref{tJEAnS_test_error_model_selection}). In Table   
 \ref{JEAnS_test_models_2} we report the average error, $\chi^2_{\mathrm{test}}$,  on unseen test data, $D_{\mathrm{train}}$ (section \ref{JEAnS_data}).   
 In all four  cases, the \jeansvii finds the true underlying models from which the synthetic data were created. 
 In Figures \ref{DLI_nat_PlumCuspOM_10k} -- \ref{DLI_nat_NonPlumCoreOM_10k} we plot the best candidate models as these were selected from the t-\textsc{JEAnS}. The plotted results were obtained after Phase III of the t-\textsc{JEAnS}. In all cases, our algorithm achieves excellent performance and reconstructs the true underlying profiles.

\subsubsection{The 1k NonPlumCoreOM dataset} 
\label{NonPlumCoreOM_1k_results}

In this section we discuss our findings for the 1k NonPlumCoreOM dataset as well as the efficiency of the GANs for synthetic data generation. The latter is  judged by the quality of the fits.  
 
The NonPlumCoreOM 1k dataset, besides being a very difficult dataset due to its strong radial anisotropy profile \citep{2017arXiv170104833R}, also presents a challenge for all Jeans moments based solvers due to its small number of data. Binning 1k data, we end up with as few as 30 binned LOS velocity dispersion values. For a small model, with only 3 knots for the definition of the B-spline basis, we end up with 5 ($\sigmar$) +5 ($\sigmat$)+ 4(smoothing penalty)+2 (DM) +2 (Stellar) = 18 unknown  parameters.  In addition, the uncertainty of the $\sigmalos$ binned values is much larger, as is evident from Fig \ref{DLI_NonPlumCoreOM_True_vs_GAN_sigmaLOS}.

For these reasons, we fit both the true 1k profile, as well as the augmented GAN profile. For the case of the 1k dataset, the test error based on the sampled mcmc values of the $\sigmalos$ bins fails to recover the correct model. 
 The augmented GAN dataset (approximately 160 binned values) selects the correct model, thus underlining the importance of this data augmentation approach.   We summarize the results of the model selection, during the EA phase, in Table \ref{JEAnS_test_models_2_1k}. 
 
\begin{figure*}
\centering
\includegraphics[width=\textwidth]{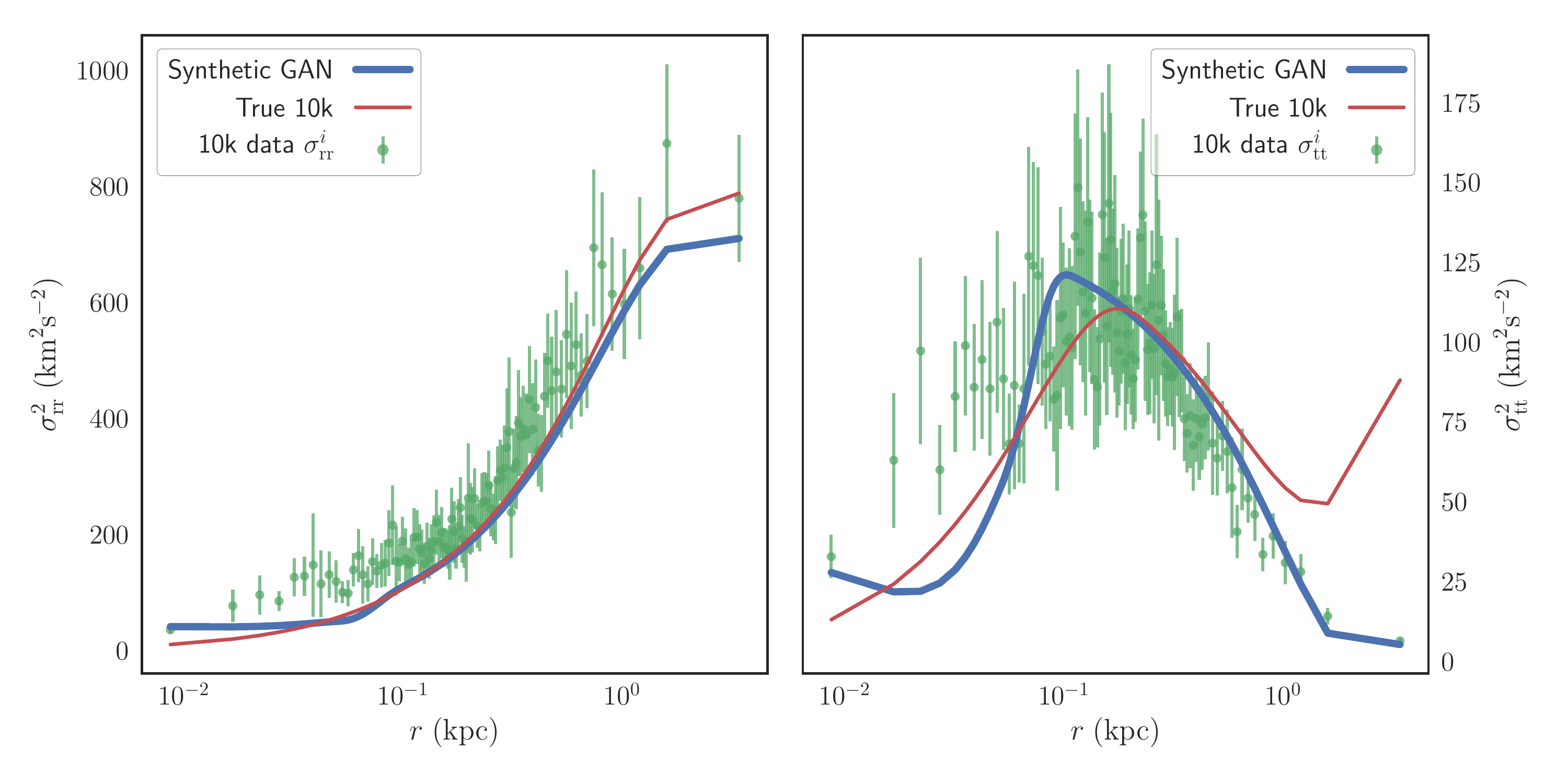}
\caption{ Difference in fitted profiles in latent profiles $\sigmar$, $\sigmat$ for the models trained on the GAN generated synthetic dataset and the 10k dataset, for the NonPlumCoreOM model. }
\label{DLI_nat_NonPlumCoreOM_GAN_vs_10k}
\end{figure*}

 \textcolor{black}{
 In Fig. \ref{DLI_nat_NonPlumCoreOM_1kGAN} we present the fit to the GAN generated data. In the top right panel the $\sigmalos$ data values are the ones created from the 25k GAN generated synthetic data. In the middle and bottom right panels, the $\sigmar$ and $\sigmat$ data values were not used in the fit. They were estimated  from the true NonPlumCoreOM 10k dataset and they are placed there for reference only. We used these because the GAN data do not have the full 3D information for us to create these data values for reference. The recovery of the dataset is much better than what we would get by using only the NonPlumCoreOM 1k dataset. The recovered profile is of lower uncertainty than the one with the NonPlumCoreOM 10k dataset, especially close to the outer regions of the data.  That is, the GAN generated dataset \emph{gives a better fit than the original True NonPlumCoreOM 10k dataset} (note that a different range is displayed on the vertical axis in all right panels of Figures \ref{DLI_nat_NonPlumCoreOM_10k} and \ref{DLI_nat_NonPlumCoreOM_1kGAN}).  This can be quantified, as can be seen in Fig. \ref{DLI_nat_NonPlumCoreOM_GAN_vs_10k}: in the left panel we plot the true $\sigma_{\mathrm{rr}}^i$ profile as this is estimated from the 10k NonPlumCoreOM dataset, as well as the highest likelihood fitted profiles, for the GAN data and the 10k NonPlumCoreOM datasets. In the right panel we do the same for the tangential dispersion, $\sigmat$. In order to quantify the quality of the fits in the unseen latent space of radial and tangential dispersions we estimate the mean square error for the radial and tangential profiles, between the best fitted profiles and the data: 
 \begin{align*}
 \chi^2_{\mathrm{rr},10k} &= \frac{1}{N_{\mathrm{bin}}}\sum_{i=1}^{N_{\mathrm{bin}}} \left([\sigmar_{\mathrm{10k}}(r_i)-\sigma_{\mathrm{r}}^i]/\delta (\sigma_{\mathrm{r}}^i) \right)^2\\
\chi^2_{\mathrm{rr},\mathrm{GAN}} &= \frac{1}{N_{\mathrm{bin}}}\sum_{i=1}^{N_{\mathrm{bin}}} \left([\sigmar_{\mathrm{GAN}}(r_i)-\sigma_{\mathrm{r}}^i]/\delta (\sigma_{\mathrm{r}}^i \right)^2 
\end{align*}
 and similarly for the tangential profile, $\sigmat$. We find for the ratios: 
 \begin{align*}
 \chi^2_{\mathrm{rr},10k}/\chi^2_{\mathrm{rr},\mathrm{GAN}} &= 0.99,  &
   \chi^2_{\mathrm{tt},10k}/\chi^2_{\mathrm{tt},\mathrm{GAN}} &= 54.26
 \end{align*}
Therefore, the quality of the $\sigmar$ fit is similar if we train \jeansvii with either the 10k dataset, or the GAN generated synthetic data. However, the quality of the $\sigmat$ fit is much worse when \jeansvii is trained with moments from the  true 10k dataset.  
This should not come as a surprise. What this means, is that from the 1k of data, the GAN system manages to recover more information than what is hidden in the moments of a 10k dataset. Then, with $\sim$160 binned $\sigmalos$ data points, it passed more information to the \jeansvii solver, than the moments of the 10k dataset can. }

\begin{table}
\caption{\textcolor{black}{Competing mass models for the NonPlumCoreOM 1k data set, with (GAN) and without (1k) data augmentation. We report the  average error on unseen test data, $D_{\mathrm{test}}$. The  true models from which the data were produced are with  {\bf bold} fonts. The lower test error is also designated with a {\bf bold} font. }}
\label{JEAnS_test_models_2_1k}
\begin{center}
\begin{tabular}{ | l |  c | c | c | c | }\hline
DataSet   & Stellar & DM      & $\chi^2_{\mathrm{test}}$ \\ \hline \hline
NonPlumCoreOM, 1k  & Plummer & NFW  & {\bf 67.4542 }   \\
 NonPlumCoreOM, 1k & {\bf  gH} & {\bf  gH }   &  69.8274  \\
\hline\hline
NonPlumCoreOM, GAN  & Plummer & NFW  & 497.068  \\
 NonPlumCoreOM, GAN & {\bf  gH} & {\bf  gH }   & {\bf 492.709} \\
\hline\hline
\end{tabular}
\end{center}
\end{table}

\smallskip
The principal criticism that is levelled at the Jeans approach is that one may find solutions to the Jeans equations that require a distribution function that is not positive at all phase-space locations, and is hence unphysical. However, one can always check that the results of our algorithm give a positive DF by testing the solution with a {\it single} Schwarzschild model. Since the solutions presented above recover the correct input dynamical models from the \GAIA, this step is not necessary here.

\section{ Discussion}
\label{tJEAnS_discussion}

We suspect that the astronomy community's definition of the Jeans degeneracy would be: \emph{many choices of functional forms for $M(<r)$ and $\beta(r)$ (or equivalently $\sigmar$) result in a $\sigmalos$ profile that is {\bf arbitrarily close} to the data. Therefore it is not possible to derive a unique mass and anisotropy profile.} This is indeed the case, the system of equations is not closed (we need additional constraints that we do not have). However, when it comes to statistical model selection the situation is  different. We can provide the additional necessary condition that closes the system of equations by selecting the ``simplest'' solution that describes well the observable data. 
The point of emphasis above in bold, that the profile should be \emph{arbitrarily close} to the data, resembles a $\chi^2$ ``selection'' criterion, which is, however, not a proper model selection method.  
The key point in \jeansvii to \emph{statistically} break the degeneracy is the realization that we can use hierarchical\footnote{With the term hierarchical we mean models that result from the same general equation, but with possibly different complexity. Examples of hierarchical models are a Fourier expansion of a function: $f(x) = \sum_{i=1}^{n} c_n, \cos(n x)$, or a B-spline basis, $f(x) = \sum_{i=1}^n c_n B_n(x)$. As $n$ increases we get models of increasing complexity that are derived from the same general equation.} models that eventually result in different quantitative fits to the data (i.e. different test error).
In other words, different assumptions of functional forms for mass, $M(r)$, and anisotropy, $\beta$, are no longer quantitatively equivalent. 

\textcolor{black}{A special note needs to be made about the fact that the notion of the mass anisotropy degeneracy, when it comes to statistical fitting, is reinforced by the fact that for the majority of stellar systems, the observables are few in number. This makes model selection even more difficult and sustains the belief 
that, given the availability of  data, it is not always possible to discriminate between competing mass models. This is more evident for moment-based mass estimators that rely on summary statistics of the initial dataset. The modern semi-supervised machine learning techniques that are actively  being developed by the community, such as the Generative Adversarial Networks for synthetic data generation, are a remedy to this problem. }

Some of the main differences of the \jeansvii that allow more efficient treatment of the degeneracy problem, assuming sufficient available data, in comparison with other approaches are: 
\begin{enumerate}
\item We do not assume two unknown parametric functional forms for both the DM mass density profile, $\dDM(r)$, and the anisotropy profile, $\beta(r)$.  This reduces the uncertainty of the parameters and allows for more robust model selection.  
\item With our choice of hierarchical  parametric models (B-splines) for $\sigmar$ and $\sigmat$ we can better statistically discriminate between competing models. This is achieved because hierarchical models find a trade-off between test and train error, and are thus more resilient to over-fitting. 
\item We incorporate a set of physically plausible constraints (Section \ref{DLI_DMS_sect}) that further reduces the feasible solution space and pushes to the limit the model selection process. 
\item For the moments solver (DMS, Section \ref{DLI_DMS_sect}), we are not using a single $\sigmalos$ value for each bin. In contrast, we are using the full MCMC chains to get additional information from the binning scheme. This allows for the estimation of train, validation and test error, as it is used in  modern machine learning supervised training techniques. 
\end{enumerate}

Although we have not performed a detailed numerical comparison by switching on and off all the constraints we used, we have the following understanding of the effect of each as well as our modelling approach: 
 \begin{enumerate}
  \item The choice of the assumed DM mass model: as with all model selection processes, our effort relies on the assumption that, if we try a large set of competing mass models, then one (or some) of them will not be very far from the truth. Then our best solution should approximate reality at a satisfactory level. 
Our contribution is demonstrating that with the use of a hierarchical basis, satisfactory model selection is possible. Obviously, if our mass model assumptions are away from the truth, we expect that the kinematic fits will also be away from the true anisotropy profile. 
  In our numerical experiments,  even with different mass model assumptions, the kinematic profiles tend to be similar. However we cannot conclude, due to the limited number of mass models and data sets we tried, that this is a general feature. In addition, we cannot quantify the ``anisotropy similarity'' in terms of similarity between competing DM mass models. This is something that requires further investigation. We also note that we have found that the choice of tracer profile affects significantly the derived anisotropy profile.  
 \item The boundary condition at the origin $\sigmar(r=0) = \sigmat(r=0)/2$ can result as a limiting case of the Jeans Eq. \eqref{eq_Jeans_trad}, as $r \to 0$, for non divergent DM potentials. However, it helps numerically inside the solver to keep it separate.  The boundary condition at the virial radius was used mainly for the domain of definition of the B-spline basis (it requires a closed finite interval). As we cannot deduce the profile further than the last datum, this constraint contributes in combination with the projected virial theorem.  
 \item The MAMMPOST-style LOSVD helps to constrain more robustly  the kinematic profile beyond the half light radius. It proved helpful in the case of the difficult NonPlumCoreOM dataset. In the other three datasets, even without it, the recovered fits were excellent. 
 \item The projected virial theorem can alter the solution space significantly, for a given mass model assumption. For example we find that if we run an MCMC exploration with and without it the parameter chains for the same model converge at different non-overlapping regions. It should also be noted that this is a very difficult constraint to implement numerically in an MCMC scheme, because it is a hard bound and does not allow efficient mixing of the chains. It is possible that there is a connection between the projected virial theorem constraint, and the approach of the virial shape parameters taken by \citet[][see also \citealt{2014MNRAS.441.1584R}]{2017arXiv170104833R}, however we have not verified this.     
 It is also interesting to note that despite the fact that the kinematic profile is essentially ``free'' after the last datum, the virial theorem still helps reducing the feasible solution space. 
\end{enumerate}

 A special note needs to be made on the particular choice of representation: in \jeansvii we represent the kinematic profile with the variables $\sigmar$ and $\sigmat$ instead of $\sigmar$ and $\beta$. This is because in the former representation, with the use of B-splines, we can linearize the system of equations (thereby greatly simplifying the solution). In contrast if we  use  $\sigmar$ and $\beta$, then from Equations \eqref{eq_Jeans_trad} and \eqref{eq_sigmaLOS_trad} it is apparent that due to the product term, $\sigmar \beta$, the system of equations is not linear. It should be emphasized that the choice of representation on its own is not adequate to statistically break the degeneracy. By linearizing the system of equations, however, we gained additional insight to the problem. The linearized equations were the key ingredient that led us to seek additional constraints (e.g. virial theorem) that further reduce the feasible solution space. 
 
 Finally, we need to emphasize again the importance of using large datasets for model selection. When these are not available, GANs can be one starting point towards the correct solution. \jeansvii --- or any other algorithm --- will fail in the absence of sufficient data.
 
\subsection{The case of multiple stellar population dynamics} 
The linearization of the system of equations in the Jeans formalism yields some useful insights for the case of multiple stellar populations. 
 When it is feasible to separate the stellar population into multiple stellar sub-populations (assuming two for simplicity) that are evolving under the influence of the same DM potential, the system of equations describing the system becomes: 
\begin{align*}
\sigma_{1\mathrm{los}}^2  &= \frac{1}{\Sigma_{1\star}(R)} \biggl(
\int_{R}^{\rvir} \rho_{1\star} K_1 \sigma^2_{1\mathrm{rr}} \diff r + 
\int_{R}^{\rvir} \rho_{1\star} K_2 \sigma^2_{1\mathrm{tt}} \diff r \biggr)\\
\sigma_{2\mathrm{los}}^2  &= \frac{1}{\Sigma_{2\star}(R)} \biggl(
\int_{R}^{\rvir} \rho_{2\star} K_1 \sigma^2_{2\mathrm{rr}} \diff r + 
\int_{R}^{\rvir} \rho_{2\star} K_2 \sigma^2_{2\mathrm{tt}} \diff r \biggr)
\end{align*} 
and the corresponding Jeans equations are:
\begin{align*}
-\rho_{1\star}\frac{\diff \Phi}{\diff r}  &= 
 \frac{\diff (\rho_{1\star} \sigma_{1\mathrm{rr}}^2 )}{\diff r} 
+\rho_{1\star} \frac{(2 \sigma_{1\mathrm{rr}}^2 - \sigma_{1\mathrm{tt}}^2)}{r}
\\
-\rho_{1\star}\frac{\diff \Phi}{\diff r}  &= 
 \frac{\diff (\rho_{2\star} \sigma_{2\mathrm{rr}}^2 )}{\diff r} 
+\rho_{2\star} \frac{(2 \sigma_{2\mathrm{rr}}^2 - \sigma_{2\mathrm{tt}}^2)}{r}.
\end{align*} 
 This is a set of four equations, with five unknowns (assuming, for simplicity, that the stellar tracer densities, $\rho_{(1,2)\star}$,  are known), namely, $ \sigma_{(1,2)\mathrm{rr,tt}}^2,\dDM(r)$. The system of equations is still not closed (in fact, irrespective of the number of sub-populations, we will always have one more unknown function than equations). However, from the insight we get from the linearized equations (say, using B-splines), we understand that if the profiles of the stellar populations are significantly different (i.e. the determinant of the linearized system is not zero), then the solution space is reduced significantly. Depending on the statistical uncertainty of the observables, this may be enough to accurately describe the underlying DM structure. In contrast, if the profiles of the sub-populations are identical, the linear systems are identical (their determinant is zero) and no additional reduction of the feasible solution space is possible.  
  Clearly, the linearization of the equations with the use of B-splines (or other suitable complete bases, e.g. wavelets), besides being a useful numerical scheme, also allows us to gain further insight into the degeneracy problem.  

For systems with multiple stellar populations where we are trying to deduce more than one kinematic profile from scarce data, the GAN synthetic data generation can be a game changer for the estimation of the different brightness and  LOS velocity dispersion profiles. The reason being that it can construct robust velocity dispersion data with small uncertainties over the extent of the system under investigation.

\section{Conclusions} 
\label{section_conclusion}

\textcolor{black}{
In this work we describe a new method for reliable mass determination independent of the mass-velocity anisotropy degeneracy. The efficiency of our method is tested on synthetic data from the \GAIA suite of mock simulations. In all cases our algorithm  reconstructs accurately the underlying kinematic profile as well as the mass content of the datasets.  
Our method includes: a) a new way of solving numerically the Jeans equations, subject to physically plausible local and global constraints, using quadratic programming. b) a new way for performing supervised learning in the framework of Jeans mass modelling, using samples from line-of-sight velocity dispersion 
MCMC chains as ``unseen'' validation and test  data sets. 
Based on this, we present  a new approach in performing regularization and  model selection. c) The application of Generative Adversarial Networks for augmenting datasets, thereby making the \jeansvii moments solver method reliable in situations where the available samples possess a relatively small number of stars.  }

\section*{Acknowledgments}
FID, GFL and CP acknowledge
support from Australian Research Council Discovery Project (DP140100198). FID also thanks the University of Western Australia for support through a Research Collaboration Awards (PG12105204). 
GFL thanks the European Southern Observatory (ESO) for support as a visiting astronomer and for hosting him in Garching where the final stages of the preparation of this publication were undertaken. RAI gratefully acknowledges support from a ``Programme National Cosmologie et Galaxies'' grant.

%
\bibliographystyle{mn2e} 
\bibliography{DLI_tgJm.bib} 

\appendix

\section{Proof of Theorem 1}
\label{Theorem_1_proof}
Let us assume that there exist two radial profiles, $\sigma_{\mathrm{1rr}}^2$ and $\sigma_{\mathrm{2rr}}^2$ that give the same LOS dispersion profile. Then: 
\begin{align*}
\sigmalos(R) &= \frac{2}{\pdtracer(R)} \int_R^{\rvir} \biggl[
K_A \biggl( \frac{\diff (\rho_{\star} \sigma_{\mathrm{1rr}}^2)}{\diff r} + \rho_{\star} \frac{\diff \Phi}{\diff r} \biggr)\\  
&\phantom{\frac{2}{\pdtracer(R)} \int_R^{\rvir} 
K_A  \biggl( \frac{\diff (\rho_{\star} \sigma_{1rr}^2)}{\diff r} + \biggr)  }
+K_B 
\rho_{\star} \sigma_{\mathrm{1rr}}^2 \biggr] \diff r \\
\sigmalos(R) &= \frac{2}{\pdtracer(R)} \int_R^{\rvir} \biggl[ 
K_A \biggl( \frac{\diff (\rho_{\star} \sigma_{\mathrm{2rr}}^2)}{\diff r} + \rho_{\star} \frac{\diff \Phi}{\diff r} \biggr) \\
& \phantom{\frac{2}{\pdtracer(R)} \int_R^{\rvir} 
K_A  \biggl( \frac{\diff (\rho_{\star} \sigma_{\mathrm{1rr}}^2)}{\diff r} + \biggr)  }
 +K_B 
\rho_{\star} \sigma_{\mathrm{2rr}}^2 \biggr] \diff r 
\end{align*}
Subtracting the above equations, yields: 
\begin{equation}
\int_R^{\rvir} \biggl[
K_A \biggl( \frac{\diff \rho_{\star}   \Delta \sigmar}{\diff r}  \biggr)  +K_B 
\rho_{\star} \Delta\sigmar \biggr] \diff r = 0
\end{equation}
where $\Delta \sigmar = \sigma_{\mathrm{2rr}}^2 - \sigma_{\mathrm{1rr}}^2$. In order for this integral to be identically zero for all values of the parameter $R$, the integrand must be zero, i.e. 
\begin{equation}
\label{ref_respo_ode_f}
K_A \biggl( \frac{d f}{d r} \biggr)  +K_B
f  = 0
\end{equation}
where we have set $f(r) = \rho_{\star} \Delta 
\sigmar$. 
For the case of $R=0$ the result is trivial $f=0$, i.e. $\sigma_{\mathrm{1rr}} = \sigma_{\mathrm{2rr}}$. For the case $r>R >0$, we manipulate Eq \eqref{ref_respo_ode_f}: 
\begin{align*}
df/dr + K_B/K_A f &= 0 \rightarrow\\
df/dr + \frac{2 r}{R^2} f = 0 \rightarrow \\
df / f = - 2r/R^2 dr \rightarrow \\
f = A \exp\{-r^2/R^2\}
 \end{align*}
where $A$ is the constant of integration, that will be determined from the virial boundary condition: since the last equation holds for all $r,R$, it will also hold for $r=\rvir$ and $R=\rvir/2$, where $\rvir$ is the virial radius of the system. However for $r=r_{vir}$, it is
\[
\lim_{r\to r_{vir}} \sigmar (r) = \lim_{r\to r_{vir}} \sigmat(r) = 0
\]
Hence, $\lim_{r\to r_{vir}} \sigma_{1rr}^2(r) = \lim_{r\to r_{vir}} \sigma_{2rr}^2(r) = 0$, i.e. $f(r_{vir})=0$. 
Then $A \exp (- 4)=0$, i.e. $A=0$, then $ f(r)=0$  and $ \sigma_{\mathrm{1rr}} = \sigma_{\mathrm{2rr}}$ for all $r$. 

This proof is also valid for  spherically symmetric systems subject to an external gravitational field: in this case  as $r\to \rvir$ both the radial and tangential velocity dispersions approach the same constant value \citep{1992ApJ...391..531D}, thus again at the virial radius of the system $f\to 0$. 

\bsp

\label{lastpage}

\end{document}